*The Photonic Lantern*

T. A. Birks*, I. Gris-Sánchez, S. Yerolatsitis
Department of Physics, University of Bath, Claverton Down, Bath BA2 7AY, United Kingdom

S. G. Leon-Saval
Institute of Photonics and Optical Science, University of Sydney, NSW 2006, Sydney, Australia

R. R. Thomson
Institute of Photonics and Quantum Sciences, Heriot-Watt University, Edinburgh, EH14 4AS, United Kingdom

*t.a.birks@bath.ac.uk

*Abstract*

Photonic lanterns are made by adiabatically merging several single-mode cores into one multimode core. They provide low-loss interfaces between single-mode and multimode systems where the precise optical mapping between cores and individual modes is unimportant.

*1. Introduction*

The photonic lantern is a low-loss optical waveguide device that connects one multimode core to several cores that each supports fewer modes [1-79]. The latter are usually, but not necessarily, single-mode. Fig. 1 is a schematic diagram of a photonic lantern made by tapering a bundle of *N* separate single-mode fibres [3]; other types differ only in detail. One end of the device is a relatively-large multimode core, and the other end is an array of several relatively-small single-mode cores. In between is a transition region, along which the waveguide changes smoothly from one type to the other. Light passing through the device will follow the transition if it is gradual enough; that is, if the transformation between the different waveguide systems occurs over a long-enough distance. This allows the photonic lantern to have low loss of light. The device is reciprocal, in that either end can act as the input.

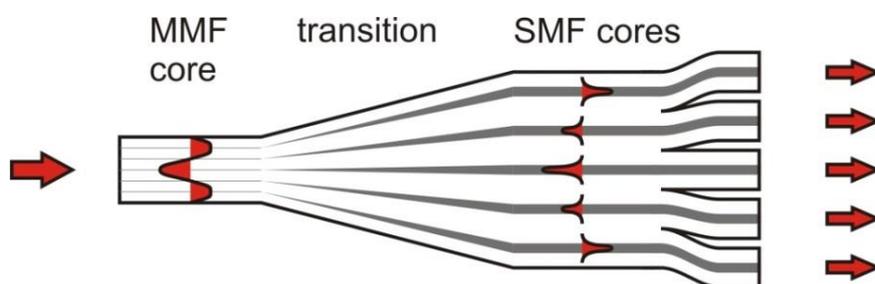

*Fig. 1. A schematic diagram of a photonic lantern made by tapering a bundle of single-mode fibres [3]. From right to left: the separate single-mode fibres (SMFs) fused together to form a single glass body, which simultaneously reduces in cross-sectional scale to form a multimode fibre (MMF) core. Surrounding it is a low-index jacket (not shown) that forms a cladding for this core.*

Note that there is no requirement for a one-to-one correspondence between the single-mode cores and the modes of the multimode core. In general, light in one core (or mode) at the input becomes distributed across most or all of the modes (or cores) at the output. Thus photonic lanterns have applications wherever there is a need to convert light between single-mode and multimode systems, especially if the distribution of light between output states (cores or modes) is unimportant. One example is in astronomical instrumentation, where the multimode light delivered by a telescope needs to be spectrally-filtered by single-mode fibre Bragg gratings [1,3,10-12,18,21,23,25,26,28,31,37,39,41,49,54,64,66]. Another is in optical telecommunications, where individual channels of information delivered by single-mode fibres need to be multiplexed via orthogonal combinations of modes in a common few-mode fibre [20,33,35,36,40,42,43,46-48,52,53,57,58,67,69-79].

Ten years after the first reports of the device [1-3], this paper reviews how photonic lanterns are made, how they work and what they are (or could be) used for. In Section 2 we describe the different techniques that have been used to make photonic lanterns. In Section 3 we look at the physics of how they work, and the design principles that should be understood when making low-loss lanterns. In Section 4 we summarise some of their potential applications. Section 5 is a brief account of the origins and usage of the term "photonic lantern". We have tried to make this review comprehensive, but our choices of topics and emphasis will inevitably reflect our own viewpoints and biases. We refer the reader to [51] for a review of photonic lanterns with a different emphasis.

## 2. How photonic lanterns are made

To make a photonic lantern we need a process for merging several single-mode cores into one multimode (MM) core, or (equivalently) splitting one multimode core into several single-mode (SM) cores. To date this has usually been achieved in two broad ways: by post-processing several single-mode fibre (SMF) cores to form a multimode fibre (MMF) core, with some kind of low-index jacket providing the multimode cladding [1,3,5,7,8,12,15,18,21,22,25,37,47,48,52,53,58-60,76,78]; or by direct ultrafast laser inscription of a pattern of waveguides into an integrated-optic chip [4,13,17,19,24,27,30,35,38,40,50,54,56,57,62,63,71,74,79]. In Section 2.1 we review the fibre tapering process, which is the key process by which most fibre lanterns have been made. Then in Section 2.2 we describe the five different types of photonic lanterns that have been demonstrated to date.

### 2.1. Tapered fibres

#### 2.1.1. The fibre tapering process

By "fibre tapering" we mean the heating and permanent stretching of one or more pre-existing optical fibres to narrow them down over a defined short distance [81-102]. This is like a scaled-down version of the process by which fibres are themselves drawn from centimetre-scale preforms. Indeed, the first photonic lanterns were tapered on a fibre drawing tower [1,3]. However, tapering is more usually carried out on a smaller scale, typically starting with a glass body that is a few hundred microns across and drawing it down to a reduced size over a centimetre length scale.

Fig. 2 shows the tapering process, together with the structure that results when a single fibre is tapered. In the middle is a narrow "waist" region, located where the heat was applied. This waist is connected to untapered fibre (which was never heated) by taper transitions at both ends. These transitions are formed from glass that was in the hot zone at the start of the process but which was pulled out of it before the end. We can call the first transition the down-taper and the second the up-taper, though this clearly depends on which way we choose to send the light.

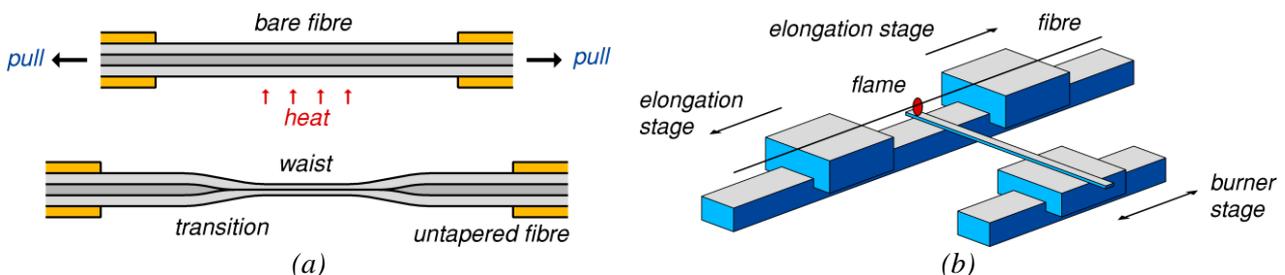

*Fig. 2. Schematic diagrams of (a) taper formation by heating and stretching a fibre and (b) a taper rig that stretches a fibre between two elongation stages while it is heated by a travelling burner.*

It is of course possible to simultaneously taper more than one fibre in this way. Indeed, the most technologically-important tapered fibre device is the four-port fused coupler, a fibre-optic beamsplitter made by tapering a parallel pair of fibres [81-87]. In that case, not only do the fibres reduce in size along the transitions but they also fuse together to form a single monolithic structure. The extent to which surface tension circularises the fused structure depends on the duration and temperature of the process [83].

The heat source must soften the glass enough for the fibre to be stretched. In the case of fibres made from fused silica (from which all fibre lanterns to date have been made), the glass temperature must be ~1700 °C or more. The most convenient heat source that is hot enough is a small flame [84-90]. In our laboratory we use a burner fuelled with butane and oxygen. Alternatives include indirectly-heated micro-furnaces [91], electrical resistance elements [84], electric arcs [92] and carbon dioxide laser beams [93]. Obviously the fibre's polymer coating must be removed from any part of the fibre placed in or near the hot zone.

Among other things, the dimensions of the final tapered fibre structure depend on the nature of the heat source. A non-uniform static heat source usually provides a non-uniform waist with a diameter that can be predicted only by calibration [84]. However, this problem is easily solved by the so-called flame brush technique, in which a small heat source (which need not be a flame) is scanned repeatedly at constant speed along along the fibre [86,89]. This yields a uniform waist with a length matching the distance scanned and a diameter that is predictable from the elongation and scan speeds. Furthermore, any predefined length and shape of taper transition can be formed if the distance scanned is appropriately varied during the process [89]. The entire set-up, including the heat source and the machinery to scan it along the fibre and to stretch the fibre, is called a taper rig, Fig. 2(b).

The waist diameter may be only slightly smaller than the original glass body, or down to sub-micron scales, or anything in between. In the case of a photonic lantern the waist provides the multimode port, with a size that is usually not very different from 100 µm in diameter. The bundle of fibres that must be tapered to produce a waist of this size can be quite large, which makes demands on the size of the heat source and places a practical limit on the number of single-mode cores that some multi-fibre lanterns (types #1 and #2, see Section 2.2) can cope with.

*2.1.2. Light propagation through a tapered fibre*

Tapering a fibre can have a profound effect on the local waveguide, Fig. 3. A typical single-mode fibre designed for telecoms applications has core and cladding diameters of 9 µm and 125 µm respectively, and a numerical aperture (NA) of 0.11. Light of 1550 nm wavelength is guided in the fundamental mode in and around the core of such a fibre, Fig. 3. Although there is some diffusion of dopants within fibres heated to high temperature [85], this is usually a minor effect. We can therefore assume that tapering simply reduces the transverse scale of the fibre's refractive index distribution in proportion to its outer diameter. Tapering the fibre down with a taper ratio of more than about 4:1 therefore reduces the core diameter of the above fibre to 2 µm or less, which is too small to guide the light effectively at 1550 nm. The fundamental mode therefore becomes guided as a cladding mode, by the waveguide at the outer boundary between the cladding and the surrounding air. In effect the entire fibre has effectively become a core with the air as the cladding, the original core being no more than a small perturbation [82,94].

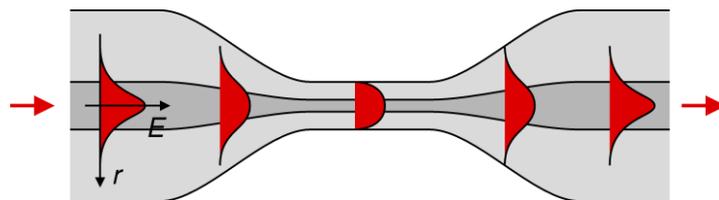

*Fig. 3. Schematic diagram of light propagation through a tapered fibre. The mode shape is shown in red as a transverse plot of field versus position. In the untapered fibre the light is localised to the fibre core. In the waist the narrowed-down core is too small to guide the light, which is instead guided by the outer cladding-air boundary.*

To get between the untapered fibre and the waist, the light must pass through the taper transition. In general, such a transition causes light to be coupled to higher-order cladding modes [94,96], which propagate as far as the point in the output untapered fibre where the fibre's polymer coating resumes. Since the coating is designed to absorb any stray light in the cladding, this light is lost. However, if the transition is gradual enough, the light in the core follows the transformation of the waveguide in a succession of local versions of the same mode [94,96]. This means that light starting in the fundamental mode of the untapered fibre steadily spreads out from the core as it shrinks, to end up in the cladding as the fundamental mode of the waist. A transition for which this is the case is called "adiabatic". The reverse process occurs in the up-taper: the light becomes recaptured by the core as it grows, leaving none in cladding modes and so avoiding loss where the coating resumes.

If the glass body being tapered includes $N > 1$ fibres, the fused waist is a single effective core (with modes) made of glass from all the fibres. To study how light from the input fibres propagates into the waist, it is helpful to consider the collection of input fibres not as individual cores with individual modes, but as a single composite waveguide with normal modes spanning all of the cores [82,83,86]. If the cores are identical and uncoupled, their individual modes are degenerate (ie they share the same propagation constant) and so can be added together arbitrarily to form valid normal modes [96].

However, at some point in the transition the cores have reduced in size and separation enough for appreciable directional coupling to occur between them. Such a system has normal modes called *supermodes*, each of which is a combination of the individual modes with definite amplitudes and phases [83,100-102]. For $N$ single-mode cores, there are also $N$ supermodes. In the case of $N = 2$, eg a four-port fused coupler, the two supermodes are the even and odd combinations of the individual modes [82,83,86]. We can therefore express the input light (in just one of the separate fibres for example) as a sum of the supermodes, and then study the propagation of the supermodes through the structure, Fig. 4. If the transition is gradual enough to be strictly adiabatic, in a generalised sense that light initially in the $n$-th normal mode remains in the $n$-th mode, then light in the $n$-th supermode of the separate cores evolves into the $n$-th mode of the taper waist, and vice versa. Provided the waist supports at least $N$ guided modes, no light is lost.

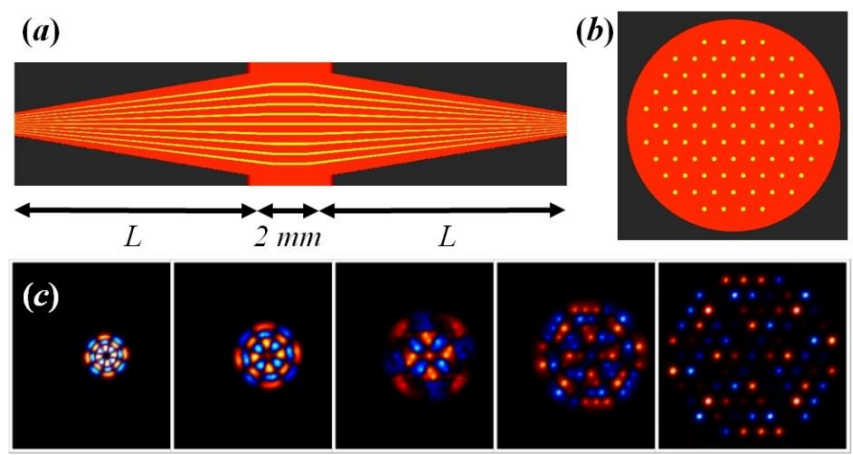

*Fig. 4. (a) Schematic side view of a multicore fibre (MCF) tapered at both ends to form an MM-SM-MM lantern pair (as discussed in Section 2.2.3). The single-mode cores are shown yellow, the fibre cladding is orange and the surrounding low-index material is black. (b) Cross-section of a model 85-core MCF. (c) Simulated propagation of an $LP_{43}$-like mode along the left half of the structure in (a) where the MCF is that in (b). The leftmost pattern is a mode of the multimode taper waist, and the rightmost image is a supermode of the array of single-mode cores. Red and blue colours represent opposite phases. Reprinted from [25].*

Here we make some (possibly) pedantic points. In fact non-adiabaticity (in the strict sense) does not necessarily cause loss. It's possible for light to couple from one supermode to another while staying in the guided set of $N$ supermodes, in which case the light is not lost. A more-relaxed practical definition of adiabaticity would therefore say that light in any of the $N$ lowest-order normal modes must remain within the $N$ lowest-order normal modes. However, it is conceptually easier to think about the stricter sense. We must also say that modes of different symmetry can change their order without implications for loss, so in the previous paragraph we should talk about the $n$-th mode *of a given symmetry*. Finally, when we count modes (eg saying that there are $N$ supermodes) we are not separately counting the two different polarisation states of the same spatial mode. This means a factor of 2 discrepancy with authors who do count them separately, though the decision to count or not count polarisations separately does not affect the design of photonic lanterns. However, our choice of counting convention does at least allow us to say that single-mode fibres guide exactly one mode.

It is often undesirable for the light field to be exposed to the air in the taper waist. This is certainly the case for photonic lanterns, where the taper waist forms the multimode port and needs to be handled or otherwise manipulated. In such cases, glass with a lower refractive index than the fibre cladding index can replace the air as the cladding material. This is most simply achieved by threading the initial fibre into a capillary of the low-index glass and tapering them all together as one glass body, the capillary collapsing onto the fused fibres by the action of surface tension [97-100].

*2.2. Five photonic lanterns*

Here we outline the five types of photonic lantern that have been reported to date [45]. Types #1 and #2 employ a collection of separate SMFs, types #3 and #5 involve single fibres with multiple cores, and type #4 is not a fibre device at all but is made from ultrafast laser-written waveguides.

*2.2.1. Lantern type #1: many single-mode fibres in a holey cladding*

We reported the first photonic lanterns in 2005 [1,3]. Their development was motivated by J. Bland-Hawthorn, a leading proponent of astrophotonics, who visited Bath in 2004 and introduced us to the OH suppression problem of ground-based astronomy (see Section 4.1). One potential solution for this problem uses aperiodic single-mode fibre Bragg gratings (FBGs) to filter the light imaged by a telescope [103]. However, even a point source (such as a star) will be imaged by a seeing-limited telescope as an extended time-dependent pattern, requiring an MMF to efficiently collect the light and deliver it to other instruments with low loss. Unfortunately, the high performance of an FBG requires it to be written in an SMF, because the modes of an MMF have a range of propagation constants and would be reflected at a corresponding range of wavelengths by a given FBG [104]. Bland-Hawthorn therefore posed the question: how can we achieve single-mode action in multimode fibre?

One of us (TAB) proposed the photonic lantern structure (and the "MM-SM-MM lantern pair" shown in Fig. 5) as a way to couple light between the MMF and the SMFs containing the gratings, and indeed to implement other types of MMF device with the performance of the SMF equivalent [1,3]. Given the expertise in our laboratory with photonic crystal fibres (PCFs) and our lack of suitable doped glass for other approaches, one of us (SGL-S) made these first lanterns by adapting the "ferrule" technique for interfacing between SMFs and PCFs [95].

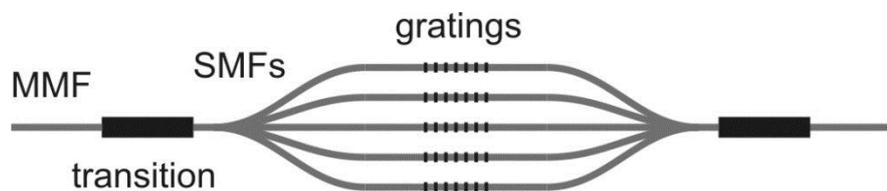

*Fig. 5. Schematic diagram of an MM-SM-MM lantern pair: two photonic lanterns connected via their single-mode fibres. In this case a multimode fibre filter is formed by including fibre Bragg gratings in the SMFs (but the structure would still be an MM-SM-SM lantern pair without them). In the experiment there were 19 SMFs. Reprinted from [3].*

PCFs are fibres made from glass with a pattern of air holes in the cladding [105]. The holes lower the effective refractive index of the cladding, allowing light to be confined to the core by total internal reflection. They are made from preforms containing holes, usually made by stacking many small capillaries with at least one solid rod to act as the core of the final fibre. In the ferrule technique the preform is a silica cane or ferrule with a hole at the core location, as well as the usual cladding holes [95]. An SMF (stripped of its polymer coating) is placed in the hole before the cane is drawn. The whole SMF then becomes part of the PCF core, so that the tapered region between the cane (with the SMF inside) and the drawn PCF provides a low-loss interface between the two types of fibre.

The first lanterns [1,3] were made using a cane with 19 holes that were just big enough to accept one SMF each, with a further ring of holes around them, Fig. 6(b). This glass body was ~3 mm in diameter, and the flame on our taper rig was far too small to deliver enough heat to stretch it. We therefore had to taper it down on our fibre drawing tower. The resulting structure had a taper transition that was a few centimetres long and ended in a length of PCF with an irregularly-shaped core incorporating glass from the $N = 19$ SMFs as well as adjacent parts of the cane, Fig. 6(a,d). This core was supported by thin glass webs in a cladding that was mostly air, Fig. 6(c). Together with the relatively large size of the core, this made the core highly multimode.

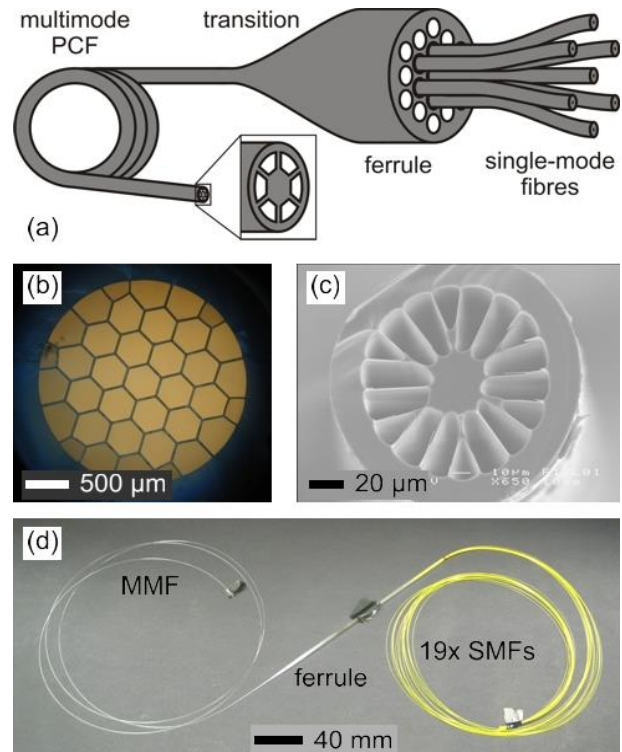

*Fig. 6. (a) A schematic diagram of the first photonic lantern, which interfaced 19 SMFs to a multimode PCF. (b) Optical micrograph of the silica cane with 19 holes, each accepting one SMF, with a further ring of empty holes around them. (c) Scanning electron micrograph (SEM) of the multimode PCF drawn from the filled ferrule. (d) Photo of the complete lantern. Reprinted from [3].*

To prove the concept of making MMF devices with the performance of SMF devices, two of these photonic lanterns were delivered to Bland-Hawthorn's laboratory for experiments with FBGs. Each of 19 identical FBGs was fusion-spliced between single-mode fibres from both lanterns, forming the MM-SM-MM lantern pair (with FBGs in between) shown in Fig. 5. Light entering the input MMF is split between the input lantern's 19 SMFs and passes into the FBGs. The filtered light continues to the output lantern, where it is funnelled into the output MMF. The unwanted light is reflected back to the input MMF. The entire structure therefore presents two MMF ports to the outside world, but internally it filters the light in a single-mode way.

The transmission spectrum of the MMF experiment is plotted in Fig. 7(a), together with the average response of the 19 SMF FBGs. Other than some noise in the MMF plot, the two spectra are indistinguishable, with a single well-defined and deep spectral notch about 0.1 nm broad. This contrasts with the spectrum that would be obtained if the same FBG had been written directly into a MMF, where the notch response would be spread over a spectrum of about 15 nm (assuming NA = 0.2) and have very little depth at any wavelength. To explore the effect of FBG uniformity, 9 of the 19 FBGs were heated by 60 °C while the other 10 remained at room temperature. The notch wavelength in an FBG's transmission spectrum is somewhat temperature-dependent, so the raised temperature gave 9 of the FBGs a slightly different notch wavelength from that of the other 10. This can be seen in the new transmission spectrum, Fig. 7(b), where there are now two notches, one for each set of FBGs (heated and unheated). Each notch is only ~3 dB deep, corresponding to the half of the light that reaches the output via the other set of FBGs. This confirmed our understanding, and underlines the importance of making sure that all of the FBGs have the same properties.

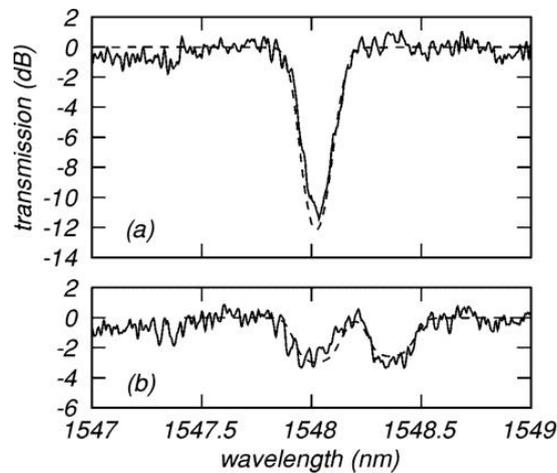

*Fig. 7. Measured transmission spectra of (solid lines) the lantern-based filter device and (broken lines) the average of the 19 SMF FBGs, when (a) all the FBGs were at the same temperature and (b) when 9 of them were heated by 60 °C. Reprinted from [3].*

Clearly there are other types of SMF devices that could be implemented like this in MMF form. It is also important to note that the function of the device does not depend in any way on how the MMF light is distributed (in either amplitude or phase) between the individual SMFs, or how the SMF light is distributed (in either amplitude or phase) between the MMF's modes. These distributions merely contribute to mode scrambling in the MMF. All that matters is whether they are low loss.

Unfortunately, in these experiments the loss was high. This is not apparent in the spectra of Fig. 7 because they are normalised to their values away from the notch. The absolute loss was 14.7 dB, which is of course far too high to be useful for astronomy. However, for these proof-of-concept experiments no attempt was made to match the number of MMF modes to the number of SMFs. As we discuss in Section 3.1, such mode-number matching is a necessary condition for low loss when the MMFs are identical and incoherently excited. In fact even this measurement was rather encouraging. From an analysis of SEM images of the MMF core, Fig. 6(c), we estimated that it supported around 710 spatial modes at the notch wavelength. If these modes were equally excited the mode-number matching principles of Section 3.1 predict a loss of 15.7 dB. The similarity of this value with the measurement suggested that it was quite possible that the poor loss was largely or entirely due to mode-number mismatch, and that more careful control of the MMF's structure could yield low-loss photonic lanterns that would be suitable for astronomy and other applications.

Further experiments were carried out in 2006 using a smaller number of SMFs and a cane with a lower air-filling fraction [106]. This allowed better control over the MMF core, since it is hard to control the very thin glass webs in a fibre like that in Fig. 6(c). One of the canes used in these experiments is shown in Fig. 8, along with the multimode PCF drawn from the cane with an SMF inserted into each of the 7 big holes. The loss measured for the lowest-loss device, for 1550 nm light coupled into each of the 7 SMFs, ranged between 0.2 and 0.5 dB with an average of 0.3 dB. This was the first low-loss photonic lantern.

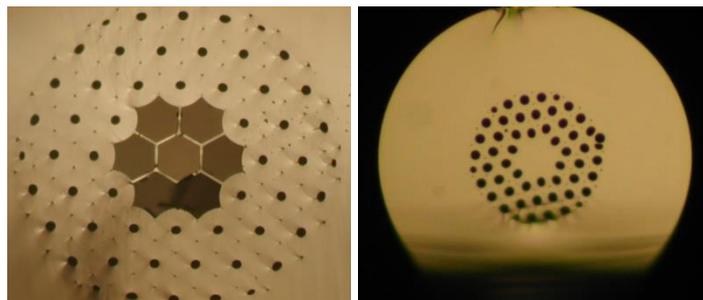

*Fig. 8. Optical micrographs of (left) a cane with 7 holes to accept SMFs and (right) a multimode PCF drawn from the cane with one SMF inserted into each of those 7 holes. The pitches of the cladding holes were 170 and 8 μm respectively.*

Although this was an improvement over the previous results and confirmed that low-loss lanterns were possible, we concluded that the PCF approach was unlikely to be satisfactory in practice. Not only did repeatable mode-number matching (and hence low loss) remain challenging, but if successful the outcome would have an MMF port that is not well-matched to commonly-used conventional MMFs. Having proved the concept, a more practical method for making photonic lanterns was needed.

*2.2.2. Lantern type #2: many single-mode fibres in a solid cladding*

The second type of photonic lantern is a variant of type #1 [5,7,8,21,22,47,48,51-53,58,65,68,77]. It was first reported by Noordegraaf et al in 2009 [5], in response to Bland-Hawthorn's quest for practical photonic lanterns capable of widespread use. As in type #1, *N* separate SMFs are inserted into a surrounding glass cane. The resulting glass body is heated and drawn down to form a taper transition to an MMF port. The MMF core is formed from the fused mass of SMFs, with the reduced-index cladding provided by the cane glass. However, now the cane is a simple cylindrical glass capillary. That is, there are no holes other than the bore of the capillary. Instead the glass in the capillary has a lower refractive index than that of the SMF cladding, thus providing a suitable cladding for an MMF, Fig. 9(a) and (b). The SMFs are bundled together inside the capillary, rather than being accommodated in separate compartments. The capillary (with the SMFs inside) is small enough to be drawn on a tapering machine (using an electrical resistance element as the heat source), which is much more practical for the purpose than a fibre drawing tower. The whole process is very similar to one used many years earlier to make multiport fused couplers [97-100], Fig. 9(c), except that the structure is cleaved at the taper waist to yield the MMF ports of a pair of lanterns.

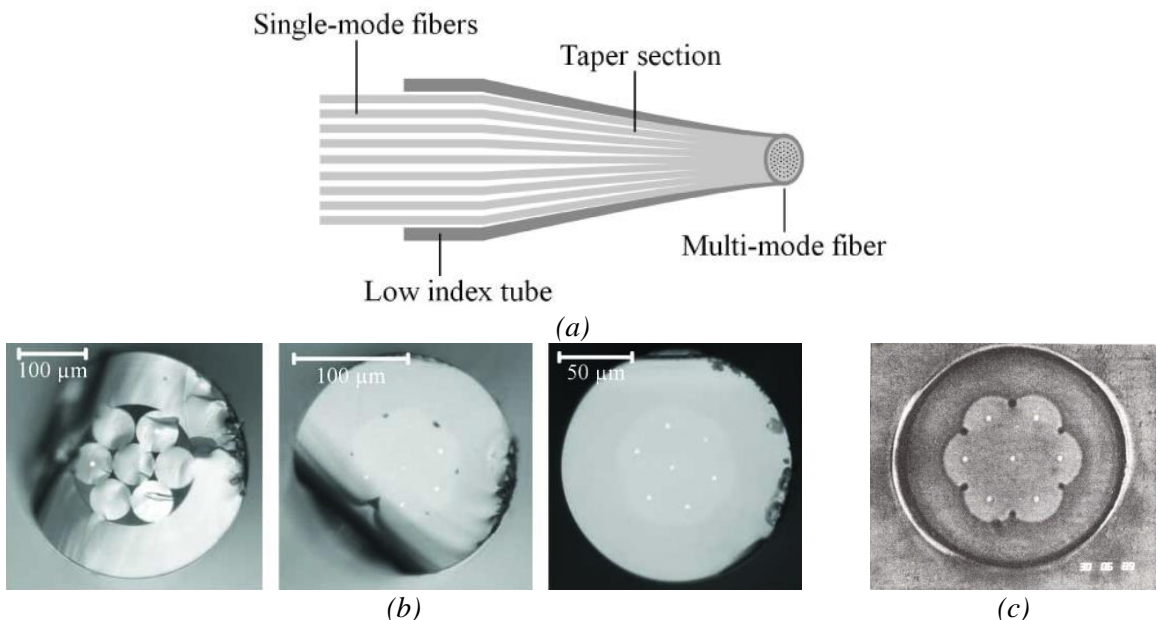

Fig. 9. (a) Schematic photonic lantern made by tapering a bundle of SMFs inside a low-index tube [8]. (b) Micrographs of the first type #2 photonic lantern, showing (left to right) cross-sections at different positions along the transition [5]. The seven SMFs and the surrounding capillary are still distinct in the leftmost image, but in the rightmost image they have fused together to form a common multimode core. (c) Cross-section of an earlier 1×7 fused coupler [100]. Reprinted with permission from D. Noordegraaf et al Opt. Express **18**, 4673 (2010) [8], D. Noordegraaf et al Opt. Express **17**, 1988 (2009) [5] and D. B. Mortimore et al Appl. Opt. **30**, 650 (1991) [100]. Copyright 2010, 2009 and 1991, Optical Society of America.

As a result, the NA of the MMF waveguide is defined by a material refractive index difference rather than by the details of a holey structure, making it much easier to control the MMF's properties than the type #1 process. The air gaps within the capillary and between the SMFs do not need to be controlled in size, they merely need to be eliminated. The resulting MMF is also closely compatible with conventional MMFs. For these reasons the type #2 process is the most well-developed method for making photonic lanterns.

For the first-reported type #2 lantern [5], *N* = 7 SMFs were bundled inside a glass capillary with an NA of 0.06 relative to undoped silica. Although the material was not reported, this NA is consistent with commercially-available fluorine- (F-) doped silica. Transmission was measured for a single device and also for two lanterns spliced at their MMF ports (the first acting as a matched source of multimode light for the second). In both cases the loss per transition was low, ~0.3 dB. In later publications the same authors describe lanterns with *N* = 61 [8] and *N* = 19 [22], Fig. 10. In both cases they measured the transmissions of MM-SM-MM lantern pairs (the configuration of Fig. 5 without FBGs) to be 70-80% over a wavelength range from 1200 nm to 1700 nm.

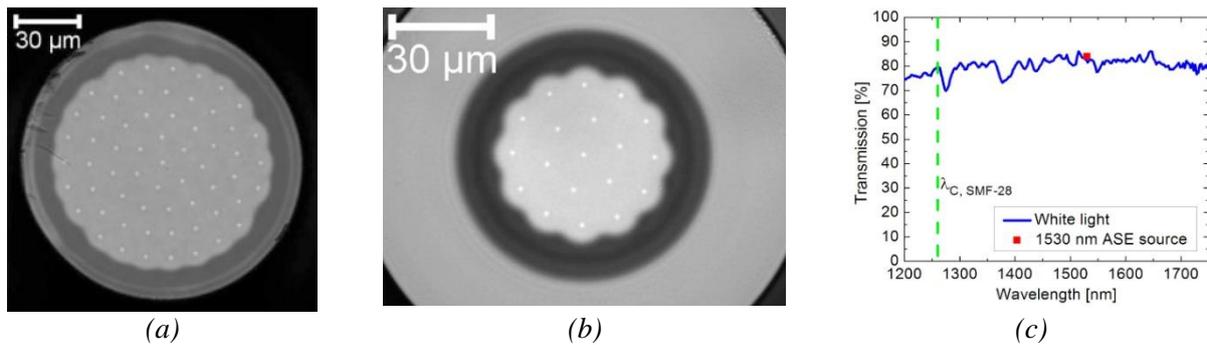

*(a)* *(b)* *(c)*

*Fig. 10. (a, b) Micrographs of (a) N = 61 and (b) N = 19 type #2 lanterns. (c) Transmission spectrum of an MM-SM-MM pair of lanterns with N = 61. Reprinted with permission from D. Noordegraaf et al Opt. Express **18**, 4673 (2010) [8] and D. Noordegraaf et al Opt. Lett. **37**, 452 (2012) [22]. Copyright 2010 and 2012, Optical Society of America.*

An important practical advance was made when a separate MMF delivery fibre was spliced onto the MMF port of a photonic lantern [22]. With the SMFs already being of indefinite length, this allows the lantern to be a completely-packaged discrete optical component that can readily be connected to conventional fibres by a typical user.

The number of individual fibres in the lantern is limited by their outer diameter relative to the capillary's inner diameter. Commercial SMFs have diameters of at least 80 μm and typically 125 μm, dictated by telecoms considerations such as the elimination of leakage and microbending losses over multi-kilometre lengths. The size of the structure being tapered can therefore become very large. For example, the MMF port of the *N* = 61 lantern of [8] was 125 μm across and had been tapered down by a factor of 10, so the initial untapered structure must have been well over 1 mm wide. The machines capable of tapering such a large glass body are specialised, expensive and temperamental. However, if the core is unchanged, reducing the SMF diameter need have little effect on fibre performance over centimetre lengths, but allows larger values of *N* for a given device size or a more compact device for a given *N*. Fig. 11 shows an *N* = 88 lantern that we made from 88 SMFs with an outer diameter of 42 μm, which compared with SMFs of 125 μm outer diameter allows nearly 10× as many cores to be fitted into the same space. It also makes it a lot easier to make the lantern adiabatic and hence low-loss [52], and allows a more-compact pseudo-slit (Section 4.3) to be formed with cores over three times closer.

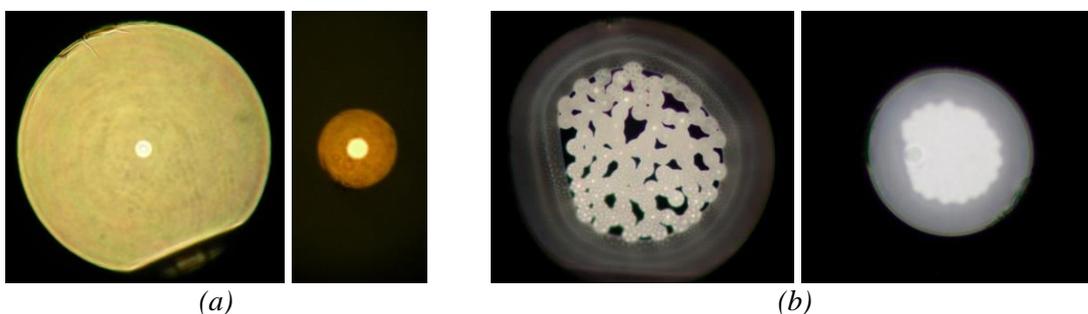

*(a)* *(b)*

*Fig. 11. (a) Micrographs, to the same scale, of SMFs with outer diameters of 125 and 42 μm but similar cores. (b) Micrographs, to the same scale, of a photonic lantern made from N = 88 of the 42 μm fibres. The SMFs and the surrounding capillary are still distinct in the left image, but in the right image they have fused together to form a common multimode core of ~60 μm diameter.*

Their low loss and repeatability, as well as being the first reliable type to be demonstrated, mean that the multi-fibre type #2 lantern is the closest to applications. Most notable among these are spectral filtering in MMF for astronomy (the original motivation) and space-division multiplexing in telecoms. These and other applications are discussed in Section 4.

*2.2.3. Lantern type #3: one multicore fibre in a solid cladding*

Although the type #2 lanterns described in the previous section have been successful, they are "hand crafted" components that are poorly scalable to large *N* [66]. The fabrication of a single pair of lanterns requires *N* SMFs to be prepared, assembled, threaded inside a tight-fitting capillary and then tapered. Then an MM-SM-MM lantern pair used to make an MMF device with SMF performance (eg, the original multimode FBG filter of Fig. 5) requires *N* identical copies of the SMF device to be obtained, and then spliced 2*N* times to the 2*N* SMF ports of the two lanterns. Furthermore, as discussed in the previous section, a bundle of SMFs inside a capillary is a relatively large glass body requiring specialised equipment to taper. All of this effort yields a single MMF device, while many astronomical observations (multi-object spectroscopy, for example) can require hundreds of MMFs.

Anticipating these problems, an alternative more-scalable solution was proposed in the original paper [3]. Instead of tapering *N* individual SMFs, we can instead taper a single fibre with *N* single-mode cores in a common cladding. This multicore fibre (MCF) needs to be fabricated specially, but several kilometres of such a fibre can be drawn at once. It can then be consumed a metre at a time, allowing one fibre fabrication effort to support the production of thousands of lanterns. As with type #2, the MCF is threaded inside a capillary of reduced-index glass before both are tapered together, but now only one fibre needs to be prepared each time. The capillary collapses onto the fibre as both are stretched to yield a MMF port, whose core is the entire narrowed MCF and whose cladding is the capillary glass, Fig. 12.

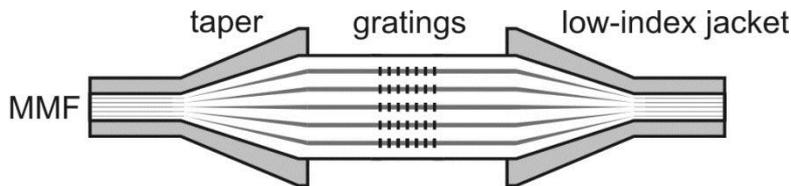

*Fig. 12. Schematic MM-SM-MM lantern pair made in tapered multicore fibre, with FBGs in the MCF's cores and a low-index jacket to provide a cladding for the MMF ports. Reprinted from [3].*

The separation of the single-mode cores prior to tapering can be much less than in the type #2 lantern, allowing many more to be packed into a given size of glass body, or more-reliable and lower-cost tapering equipment to be used for a given number of cores *N*. Furthermore, the FBGs can be written in one shot into all of the cores of the MCF at once. The number of steps to produce a complete MMF filter is therefore reduced by a factor of order *N*.

After proving the key elements of this concept in 2010 using a spare all-solid photonic bandgap fibre [12], we demonstrated MCF lanterns using a purpose-made fibre with 120 single-mode cores [18,25,37,59], Fig. 13. The fibre had an outer diameter of 230 µm, which was comfortably within the limit of what could be processed by our simple flame-based taper rig. This was tapered within an F-doped silica capillary with an NA of 0.21 to form a MMF core of 50 µm diameter. The losses of MM-SM-MM lantern pairs were <0.5 dB.

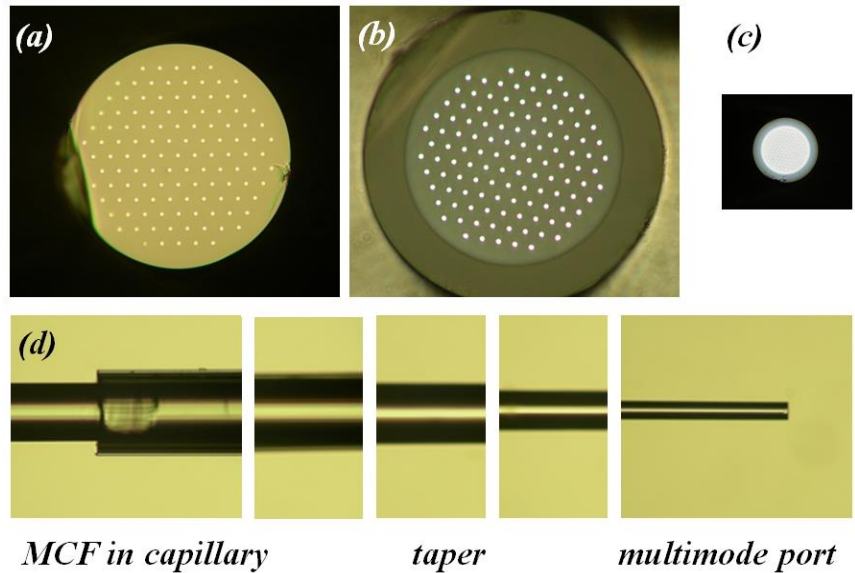

*Fig. 13. Cross-sectional micrographs, to the same scale, of (a) an N = 120 MCF, (b) the MCF with an F-doped jacket collapsed around it and (c) the jacketed MCF tapered down to form a lantern's MMF port. (d) Montage of transverse micrographs of the MCF lantern. Reprinted from [25].*

It is worth pointing out one key benefit of the multi-fibre type #2 and type #1 lanterns that type #3 lanterns do not have. The presence of *N* separated single-mode fibres, rather than *N* cores fixed within a multicore fibre, allows one to easily couple light independently into or out of each SM core, and indeed to position those cores freely in 3-D space. For example (see Section 4.3) the SMFs can be lined up to produce a pseudo-slit, perpendicular to which the light distribution is diffraction-limited, which is desirable at the input to a spectrograph [10,23]. Multicore lanterns are therefore better-suited to applications where access to individual SM cores is not required, for example for FBG filters or focal-ratio preserving multimode guides or mode scramblers [25,60]. Nevertheless individual access to the MCF's cores is possible with the aid of a fan-out device, and free-space techniques can yield the separate spectra of the cores (see Section 4.3). A suitable fan-out has been demonstrated for our 120 core MCF using ultrafast laser inscription [13,24] and, as discussed in the next section, the technique can also be used to make the lantern itself.

*2.2.4. Lantern type #4: directly-written integrated waveguide chip*

The three types of photonic lantern reviewed so far all involve the tapering of optical fibres containing single-mode cores. Where the cores reduce in size they cease to be effective waveguides. The light spreads out to fill the entire fibre or collection of fibres, which now acts as a multimode core within whatever low-index material surrounds it. However, it is also possible to conceive of a waveguide transition in which the SM cores do not shrink into insignificance, but instead are simply brought closer together until they meet to form a big multimode core. If this transition is adiabatic, it should act as a photonic lantern.

It is hard to imagine forming such transitions in (conventional) fibres using the techniques outlined above, but they have been demonstrated in integrated optic waveguides made by ultrafast laser inscription (ULI), Fig. 14 [4,17,19,27,30,35,50,71,74,107-126]. This is a relatively-new direct laser writing technique for locally modifying the structure of a substrate material in three dimensions [107,108]. A lens is used to focus ultrashort (fs or ps) pulses of sub-bandgap radiation inside the material. Because the pulses are so short, the peak electric field at the focus can be high enough to induce nonlinear absorption processes, which transfer energy from the pulse to the material at the focus. The deposited energy first induces a free-electron plasma, which transfers its energy to the material lattice on the ps timescale. This rapidly heats the lattice, can melt or even fully ionise the lattice. The lattice cools and resolidifies on the ns and μs timescale, freezing-in a variety of mico- and nano-scale structural changes. Some of the changes, such as densification [109], expansion [110] and element redistribution [111] are intuitively understandable given the extreme physical conditions; others, such as the formation of sub-wavelength nano-gratings [112], are less so.

The structural modification induced depends on the material and numerous ULI parameters like pulse energy, substrate translation speed, pulse duration and pulse focusing. ULI can locally modify the chemical etch rate [113], thermal conductivity [114] and optical absorption [115] of a range of materials. It is however the local modification of a dielectric's refractive index that has attracted the most attention, since it enables the fabrication of 3D optical waveguides simply by translating the material through the laser focus [107,116], Fig. 14(a). This capability has been intensively investigated for applications including optical waveguide amplifiers [117], waveguide lasers [118] and 3D interconnects [119], and can even be used to write optical waveguides inside optical fibres [120,121].

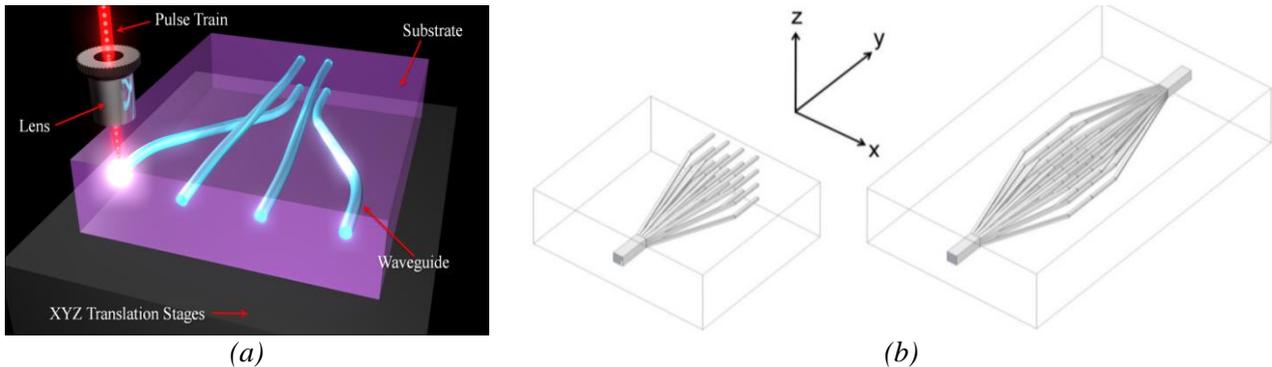

*(a)* *(b)*

*Fig. 14. Schematic diagrams of the ULI technique. (a) The inscription of an integrated optic waveguide using a focused femtosecond laser beam. (b) (left) A schematic photonic lantern in which 16 square single-mode cores meet to form a square multimode core and (right) two such lanterns forming an MM-SM-MM lantern pair. Reprinted from [17].*

In 2009, Thomson et al proposed that ULI could be used to make 3D integrated photonic lanterns [4]. Since the original motivation for photonic lanterns came from astronomy, the authors proposed that such devices could be seamlessly combined with functions such as reformatting (eg for coupling to a photonic spectrograph) and optical filtering (by integrating Bragg gratings into the waveguides).

The first ULI-fabricated photonic lanterns were demonstrated in 2011 by authors of this paper [17,19]. Each one comprised a 2D array of small SM cores that approached each other along the third dimension, to merge and form one large MM core, Fig. 14(b). The ULI parameters were set so that the cross-section of the induced refractive index change in the borosilicate glass substrate was directly determined by the beam waist and confocal parameter of the laser focus. The structure was written in the so-called transverse writing geometry, by translating the substrate perpendicularly to the propagation axis of the laser beam. Lenses with numerical apertures of ~0.5 are generally used to focus the laser beam, which results in an inscribed index profile that is much longer along the path of the beam than it is wide. To correct this asymmetry, the multiscan technique was implemented [122,123]: the individual SM cores were formed by scanning the substrate through the laser focus 20 times, each with a small transverse displacement, so that each laser-written track abuts the previous one to build the desired square core cross-section. One advantage of the multiscan technique is that it can be used to inscribe waveguides that exhibit a refractive index profile that is close to step-index, as shown in [17]. As a result, complete photonic lantern transitions were fabricated by creating transitions where the 4×4 array of individual SM cores approached each other over a length of 30 mm to form a single large MM core with a quasi-step-index square cross-section, Fig. 15.

All characterisation was performed using 1550 mm light. The measured insertion loss of an MM-SM-MM lantern pair, with an overall length of 70 mm, was 5.7 dB. The difference between this and the 5.0 dB loss of a multimode waveguide of the same length could be explained by a mode-number mismatch (see Section 3.1.2), in which case the lantern structure does not add significantly to the loss of the waveguide itself. A single lantern with no superfluous lengths of waveguide was estimated to have a loss of 2.0 dB, corresponding to an attenuation of 0.7 dB/cm. This is rather high for ULI, which can achieve attenuations down to 0.1 dB/cm for light in the 1550 nm band [123].

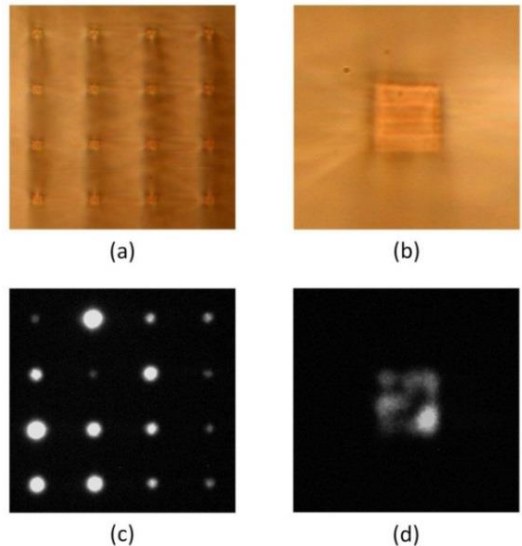

*Fig. 15. Micrographs of (a) the single-mode array of cores and (b) the multimode core in a ULI lantern, respectively. Light patterns emerging from (c) the single-mode array and (d) the multimode core, for input light in the other end. The fields of view are 200 μm (a) and (c) and 100 μm (b) and (d) across respectively. Reprinted from [17].*

An alternative ULI approach was implemented by a different team [27,30,50,54], who utilised ULI parameters (primarily a high pulse repetition rate of ~ 5 MHz) that resulted in strong thermal accumulation in the focal region. This produced waveguide cores with much more complex index profiles than those achieved using the multiscan technique, consisting of strong negative and positive refractive index changes. Although such cores cannot be neatly abutted to form a quasi-step-index MM waveguide like that of [17], the authors successfully demonstrated that a close array of 19 such cores could be brought together to form an MM core if these elements did not overlap [27]. The result was a more-circular core but with a less-uniform index distribution. The insertion loss of a single MM-SM-MM lantern pair was around 2.1 dB, and the authors determined this was due to the ~0.3 dB/cm substrate absorption for 1550 nm light. They also demonstrated a MM-to-linear-array refromatter with a loss of ~0.8 dB. In later work, the same group showed that ULI-fabricated Bragg gratings could indeed be integrated into an MM-SM-MM lantern pair [54], enabling high-resolution spectral filtering of MM light with a comparable notch depth to that achieved by the multicore fibre approach. Such devices may have applications in astronomy for OH-line suppression.

The ULI approach to making photonic lanterns has a number of advantages over fibre methods, and one serious disadvantage. Among the advantages are its great versatility. The waveguides can be directed virtually anywhere within the glass substrate, in ways that cannot be achieved within a fibre. This is illustrated by the idea of forming a lantern in which the output single-mode cores are lined-up to produce a pseudo-slit, as proposed in [4]. This can be achieved using the free SMFs in a type #2 lantern, Section 2.2.2, but the separation of the cores cannot be less than the SMF diameter. In contrast, in a ULI device the cores can be as close as desired [24] and may even merge to form a true slit-shaped core [56,63]. ULI lanterns are intrinsically stable, being monolithic glass chips, which may be important where an MM-SM-MM lantern pair acts effectively as a multipath interferometer. The technique can also be readily extended to glasses with long-wavelength transmission, allowing lanterns to be made for mid-infrared applications [62]. Finally, the technique lends itself to repeatable manufacturing. While fibre lanterns must be made one at a time, many ULI lanterns can be inscribed in one process on a single substrate. Indeed, although it appears that the early type #2 lanterns were made commercially to order [10,28,64], ULI was used to make the first photonic lanterns to be generally available commercially, through a company (Optoscribe) co-founded by one of us (RRT) [79]. ULI lanterns supplied by Optoscribe have been used by a number of groups in recent test-bed telecommunications demonstrations [35,74,80].

The serious disadvantage (at present) is of course the relatively high propagation loss compared to fibre based devices. The loss of any optical waveguide is due to scattering, radiation and absorption. The key to enabling ultra-low loss ULI devices is finding a material, with low absorption, that enables the ULI of waveguides with little scattering and a high-enough index change for waveguide bends with low radiation losses. Finding such a material is not easy, but significant progress is now being made to enable low-loss tight bends [124] and low absorption [125]. While this remains an issue, an alternative approach to maximise throughout could be to combine the fibre and ULI techniques, taking advantage of the best aspects of both [24].

Integrated optic waveguides have long been made in other ways besides ULI, and some of them have been used to demonstrate lantern-like structures [127-129]. Unlike ULI, these technologies are essentially confined to the plane. They therefore lack the ability to form paths in 3D or large cores that could interface well with multimode optical fibres. We are not aware of any current work on photonic lanterns with these planar technologies.

### 2.2.5. Lantern type #5: hole collapse in multicore PCFs

A fifth way to make photonic lanterns was demonstrated in 2013. Like lantern type #3, the substrate is once again an optical fibre with several single-mode cores. However, in this case the fibre is not tapered down to make those cores insignificant. Instead, the cladding material separating the cores is converted into core material, in effect making the cores bigger until they merge to form a multimode core. Such transitions can be made in photonic crystal fibres (PCFs) in which the cladding is defined by the presence of air holes [105]. Making the holes disappear leaves behind solid glass that matches the core material.

This is achieved simply by heating the PCF with one of the same heat sources that are used on a taper rig, but this time without significantly stretching the fibre [90]. Surface tension causes the holes to shrink and collapse completely as the hot glass flows. To allow the outer holes to stay open (so that the final multimode core has a cladding) they can be pressurised to oppose the action of surface tension. Such differential pressure can be provided by blocking (with glue) the ends of the holes that should collapse but not the ones that should stay open. Application of uniform pressure to that end of the fibre will cause only the latter holes to be pressurised, Fig. 16.

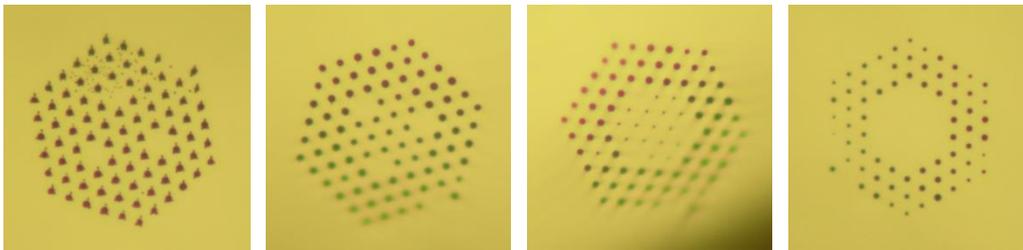

*Fig. 16. Cross-sections of a 3-core PCF (left), in which the holes between the cores have been gradually collapsed over a 4 cm length to form a single multimode core (right).*

The first-reported type #5 lanterns were made with dissimilar SM cores and acted as mode multiplexers (Section 4.4), exciting the individual modes of the MM core from light in each SM core [46,52]. However, the more-straightforward photonic lantern of Fig. 16 results if no effort is made to make the cores dissimilar, in which case each SM core excites a distribution of the modes of the MM core.

### 3. How photonic lanterns work

Light propagation through a photonic lantern closely resembles that through several fibres tapered together as described in Section 2.1.2, even for those lanterns that are not in that literal form. Light in a given mode of the multimode core excites light in the separate single-mode cores with a given distribution of amplitudes and phases, as determined by a supermode in that part of the transition where the single-mode cores are far enough apart to be distinct yet close enough to be weakly coupled [3,9,18,21,25,36,51], Fig. 17(a). In the other direction, light in a given single-mode core excites light in several modes of the multimode core, as determined by the superposition of supermodes that sum to give light in only that one single-mode core, Fig. 17(b). We can say that the array of single-mode cores is just another multimode waveguide, but its modes (ie, the supermodes) are all degenerate and so freely couple together: a photonic lantern is then simply a device to interface this multimode waveguide with a more conventional one.

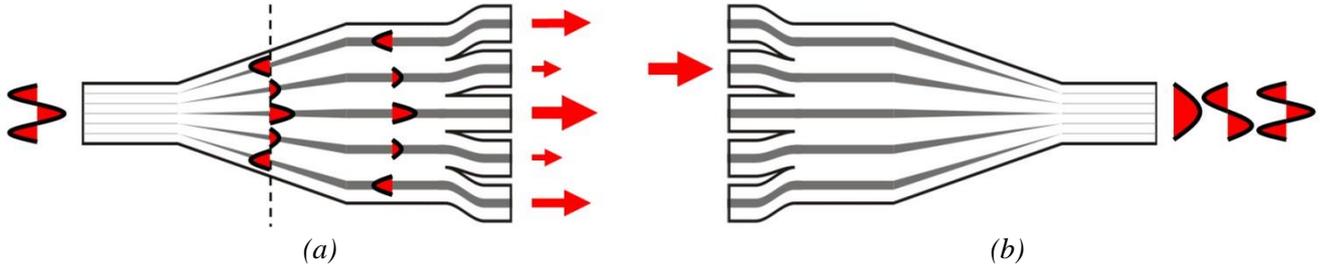

*(a)* *(b)*

*Fig. 17. Schematic diagrams of light propagation through a photonic lantern (a) from one mode of the multimode core to the array of single-mode cores and (b) from one single-mode core to the multimode core. The dashed line indicates a point in the transition where the single-mode cores are far enough apart to be individual waveguides yet close enough to be weakly coupled. This is where the supermode description of the normal modes is appropriate. Propagating towards the array of separate single-mode cores, the relative amplitudes and phases in the cores stay the same and the cores simply become uncoupled. Propagating in the other direction, each supermode evolves adiabatically into a mode of the multimode core. Because light in one core is a superposition of several supermodes, several modes of the multimode core are excited.*

Clearly we want propagation of light through the lantern to be low in loss. We will not consider here those imperfections that can be present in any waveguide, such as attenuation by the material itself or scattering by dirt and kinks due to poor manufacture. The three recognised causes of loss specific to photonic lanterns (and which can be reduced or eliminated by careful design and fabrication) are decreases in mode number, mode-order changes due to symmetry and non-adiabatic transitions. We now discuss these in turn.

*3.1. Mode number*

*3.1.1. Thermodynamics*

To simplify the analysis we assume that each mode of the multimode core evolves along the transition into a distinct supermode of the single-mode array, and vice versa, although really it is only necessary to assume unitary propagation (that the modes evolve into mutually-orthogonal states). In any case, if the light is unpredictably or incoherently distributed among the $N_{in}$ modes of the input system (the multimode core or the single-mode array) then for low loss we need

$$N_{out} \geq N_{in} \tag{1}$$

modes in the output system. Otherwise there could be some input light with no guided state to accept it at the output, which must therefore go somewhere else (ie, unguided radiation modes) and be lost.

This is an expression of the brightness theorem of optics, or indeed the second law of thermodynamics [130]. If we consider the number $N$ of unpredictably or incoherently excited modes to be analogous the number of microstates of a statistical-mechanical system, then Boltzmann's entropy $S$ for the mode distribution is

$$S = k_B \ln N \tag{2}$$

where Boltzmann's constant $k_B$ is irrelevant to the discussion but included for consistency. The same arguments that in statistical mechanics require entropy to never decrease also require the number of incoherently-excited degrees of freedom of an optical system to never decrease. If our waveguide structure does not provide enough of these guided degrees of freedom, radiating degrees of freedom (ie, loss) will make up the shortfall.

Thus a low-loss transition from an incoherently-excited multimode core to one single-mode core is unphysical. The photonic lantern avoids this prohibition by having several single-mode cores instead. The connection with thermodynamic is no mere analogy: a perpetual motion machine could be built from a collection of waveguides that funnel incoherent multimode light into a single mode with low loss!

It is important to note that the number of statistical-mechanical states equals the number of modes only if the modes are excited unpredictably or incoherently. The input light could instead be distributed across all of the modes of the MM core (say) but with definite and predictable amplitudes and phases. The light is then in a single pure state (in the quantum mechanical sense), albeit a complicated one: $N_{in} = 1$. It is then possible in principle to couple the light into a single mode with low loss, given a suitably-sophisticated coupling platform such as the mode-routing device described in [131]. The simplest example is an unrealistically-perfect MM-SM-MM lantern pair, where light in one mode of the input MM core excites a single supermode in the array of SM cores. If the SM array is perfectly straight, stable and composed of identical cores of identical length, Fig. 18(a), then the pattern of amplitudes and phases of the light in the cores is maintained along the SM section. That one supermode is therefore delivered intact to the output lantern, which returns it back to the original mode of the output MM core. This is the case even if there are many more SM cores than there are modes of the MM core.

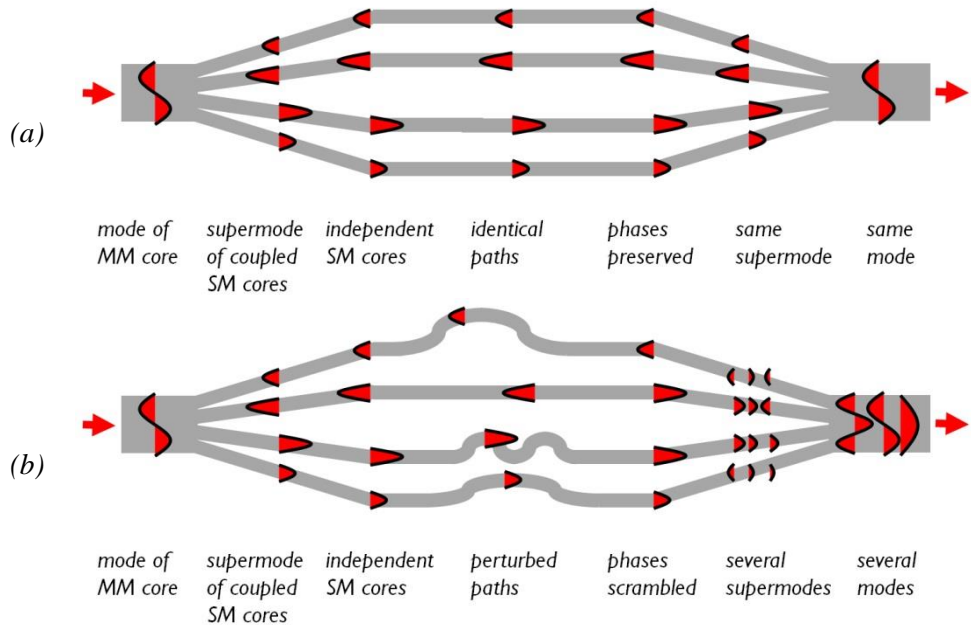

*Fig. 18. Schematic MM-SM-MM lantern pairs where the intermediate single-mode cores are (a) equal in optical path length, or (b) have unknown or uncontrolled differences in optical path length, perhaps because of different physical lengths, bends and/or differences in propagation constant due to core size and refractive index variations. The perfect case (a) preserves the relative phases in the cores so that the output is in a single supermode, whereas the more realistic case (b) scrambles the phases and so excites many other supermodes.*

Although perhaps type #4 ULI lanterns (Section 2.2.4) are compact, stable and well-defined enough to approach this behaviour, in practice an array of real SM fibre cores will not be perfect, Fig. 18(b). Independent SMFs will inevitably have different lengths, and even a multicore fibre will have accidental bends and core-to-core dissimilarities. The phase relationships between the SM cores will therefore not be maintained along the SM section, and indeed may change with time due to environmental perturbations, causing a comprehensive redistribution of the light across most if not all of the supermodes. This scrambling of the supermodes effectively "thermalises" the state of the light, ensuring that $N_{in}$ equals the number of all available modes or cores. Even if in principle the modes are coherent and could be unscrambled, in practice we can usually assume that light at any point along a lantern incoherently fills all available states.

Given the historical connection of photonic lanterns with astronomy, it is interesting to note that a photonic lantern combined with a mode-routing device (like that of [131]) is a guided-wave version of the free-space adaptive optics (AO) techniques used on ground-based telescopes. AO uses a calibration source, such as a bright star or artificial guide star, to rapidly measure the phase distortion imposed on the incoming wavefront by the atmosphere. A deformable mirror can then impart a corrective phase profile on the wavefront, so that the phases of the modes that constitute the (apparently) incoherent speckle pattern are adjusted to reconstruct a point-spread function closer to the diffraction limit. One could certainly imagine implementing this on a telescope using a photonic lantern and a mode-routing device in place of the deformable mirror [132], but it is not clear how the corrective phase information for input into the mode-routing device would be obtained, particularly in a photon-starved regime.

*3.1.2. Mode-number matching*

In the case of a fibre structure with a circular step-index multimode core of radius *a* and refractive index $n_1$ in a cladding of index $n_2$, the number *N* of guided modes for light with a wavelength of $\lambda$ is determined by the V-value

$$V = \frac{\pi d NA}{\lambda} \tag{3}$$

where the numerical aperture

$$NA = \sqrt{n_1^2 - n_2^2} \tag{4}$$

is a convenient way to represent the refractive index step.

The relationship *N(V)* between the number of modes and *V* is monotonically-increasing but complicated and step-like, as modes become guided at defined values of *V* given by the zeros of Bessel functions [96]. Our calculation of this function (for a weakly-guiding circular step-index core) is plotted in Fig. 19. However, for large values of *V* and *N* there is a very simple asymptotic expression [96]

$$N \sim \frac{V^2}{4} \tag{5a}$$

where we emphasise (Section 2.1.2) that we are not counting the two polarisation states of each spatial mode separately (otherwise the dependence is $\sim V^2/2$). The asymptotic expression is also plotted in Fig. 19, where the similarity with the exact calculation is evident. A different asymptotic formula for *N(V)*, from [133], has also been used for the design of photonic lanterns [5,8,10] and (not counting polarisation states):

$$N \sim \frac{1}{2}\left(2 + \frac{4V^2}{\pi^2}\right) \tag{5b}$$

This too is plotted in Fig. 19, showing that it is rather less accurate and so should be avoided.

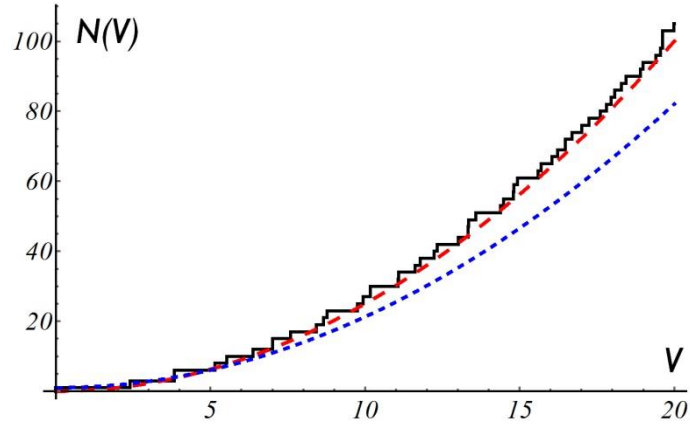

*Fig. 19. The number of modes N of a circular step-index fibre core as a function of V. The solid black stepped curve is the exact calculation based on the successive cutoff of modes, and the dashed red line is the asymptotic formula of Eq. (5a). The dotted blue line is the inaccurate formula Eq. (5b) from [133].*

Substitution of Eq. (3) into Eq. (5a) gives us a simple approximate expression for the number $N_{MM}$ of guided modes in the multimode port of a photonic lantern made by tapering fibres

$$N_{MM} \approx \left(\frac{\pi d NA}{2\lambda}\right)^2 \tag{6}$$

The number of guided supermodes of an array of $N_{SM}$ weakly-coupled single-mode cores is just equal to $N_{SM}$, again not counting polarisations separately. For the two directions of propagation through a photonic lantern, Eq. (1) therefore requires

$N_{SM} \geq N_{MM}$ (MM input to SM array output) (7a)

$N_{MM} \geq N_{SM}$ (SM array input to MM output) (7b)

Thus the simplest way to avoid losses due to mode-number mismatching can be described as the "light bucket" method, where the output of each photonic lantern in a system has a greater capacity for modes than the input, Fig. 20. (For a broadband system, this means setting $\lambda$ in Eq. (6) to the shortest wavelength at an MM input and the longest wavelength at an MM output.)

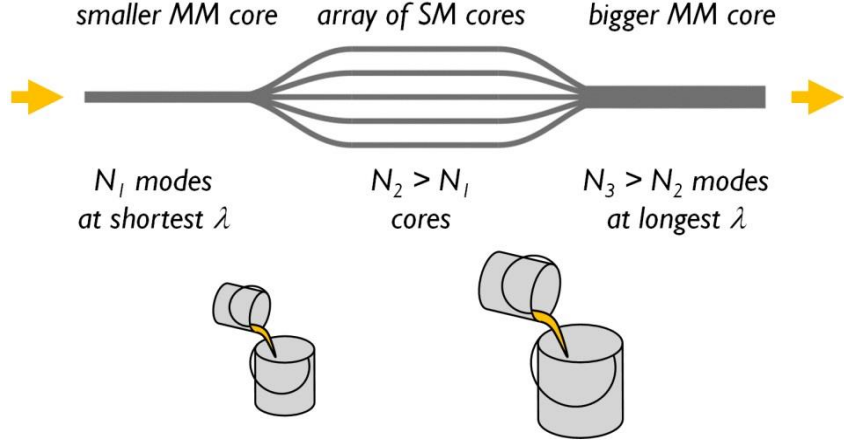

*Fig. 20. The "light bucket" method for avoiding mode-number mismatch losses, illustrated here for an MM-SM-MM lantern pair. Each section supports more modes than the previous one. Likewise it is easier to pour water from one bucket into another, without spilling any, if the second bucket is bigger.*

Such an approach is simple to implement and guarantees no mode-number mismatching losses over an arbitrarily large wavelength range (provided the SM cores remain single-mode). A small additional margin in the mode numbers at each stage also guarantees no symmetry losses (Section 3.2) as well. However, as in thermodynamics, it is not reversible. The lantern would only be low-loss in one direction, and the output of an MM-SM-MM lantern pair would need a bigger core and/or a higher NA than the input. (This argument assumes scrambled "thermalised" inputs of course, Fig. 18: the lantern pair is still a reciprocal device, not an optical isolator.) The extra modes degrade the brightness, which in some applications is undesirable. For astronomy it gives rise to focal ratio degradation (FRD, see Section 4.2), with important consequences for the design of instrumentation (such as spectrographs) downstream of the lantern pair [134].

Therefore, from now on we will assume that additional modes, brightness degradation, irreversibility and FRD are to be avoided by requiring the number of modes to be equal along the structure so that $N_{MM} = N_{SM} = N$. To yield the equalities in Eq. (7), the number of SM cores should be

$$N = \left(\frac{\pi d NA}{2\lambda}\right)^2 \quad \text{(mode-number matched)} \qquad (8)$$

This is the single most important design equation for photonic lanterns [25]. It allows us to estimate the number of single-mode cores we need to match a given multimode fibre system, or the diameter of the multimode core that matches a given number of single-mode cores. For example, for a system of $N = 120$ single-mode cores and a multimode NA of 0.215, the multimode core's diameter should be 50 μm for mode-number matching at the wavelength of 1550 nm.

The result Eq. (8) is not affected by whether polarisations are counted separately, which would merely have introduced a factor of 2 onto both sides of the equation. It does however depend on the multimode core being step-index and circular. A different relationship for $N_{MM}$ would be needed for square multimode cores made by ULI [17], though the parametric dependences are the same.

*3.1.3. Loss and its wavelength dependence*

The loss due to a decrease in mode number through a lantern ($N_{out} < N_{in}$) depends on how the input waveguide is excited. A simple calculation is possible if we assume that all the input modes are equally excited. In that case there are $N_{in}$ units of input power and $N_{out}$ units of output power, so the loss in decibels (expressed as a positive number) is

$$loss = -10\log_{10}\left(\frac{N_{out}}{N_{in}}\right) \quad \text{dB} \qquad (N_{out} \leq N_{in}) \tag{9}$$

If mode number increases ($N_{out} > N_{in}$) there is no such loss. For an MM-SM-MM lantern pair with identical input and output cores but $N_{MM} \neq N_{SM}$, the mode number could decrease either in the first lantern or the second. However, in either case the mode-mismatch loss of the lantern pair is

$$loss = 10\left|\log_{10}\left(\frac{N_{SM}}{N_{MM}}\right)\right| \quad \text{dB} \tag{10}$$

This reasoning has provided circumstantial evidence that the losses measured in early experimental MM-SM-MM lantern pairs were due mainly to mode-number mismatches rather than more fundamental problems. For example, the high losses measured for the type #1 lantern (made by PCF drawing, Section 2.2.1) could be accounted for by the estimated ratio $N_{SM}/N_{MM} = 19/710$, which from Eq. (10) gives a 15.7 dB loss comparable to the experimentally-measured loss of 14.7 dB [3]. Similarly, for the first type #4 lantern (made by ULI, Section 2.2.4) the corresponding ratio $N_{SM}/N_{MM} = 16/13$ gives a 0.9 dB loss comparable to the experimentally-measured loss of 0.7 dB [17].

The number $N_{MM}$ of modes of a multimode core depends on wavelength Eq. (6), whereas the number $N_{SM}$ of single-mode cores is fixed. Mode-number matching in a simple photonic lantern is therefore only possible at one wavelength [2,25]. The wavelength dependence of loss due to mode-number mismatch can be estimated by substituting Eq. (6) into Eq. (10) to give

$$loss = 20\left|\log_{10}\left(\frac{\lambda}{\lambda_0}\right)\right| \quad \text{dB} \tag{11}$$

where $\lambda_0$ is the matching wavelength that satisfies Eq. (8). Since $N_{MM}$ will have the same wavelength dependence for any conventional waveguide, Eq. (11) should apply even for multimode cores other than a circular step-index fibre [17], making Eq. (11) a very general result with just the one parameter $\lambda_0$. An example of this dependence is plotted in Fig. 21.

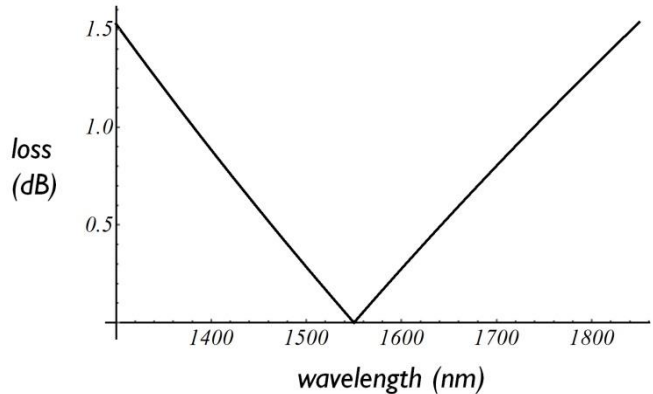

*Fig. 21. The calculated wavelength dependence, from Eq. (11), of the loss due to mode-number mismatch in an MM-SM-MM lantern pair designed to match mode numbers at a wavelength of $\lambda_0 = 1550$ nm. The second-mode cutoff wavelength of the SM cores is assumed to lie outside the wavelength range.*

Interestingly, experimental tests have reported that the loss of photonic lanterns need not depend on wavelength as strongly as Eq. (11) [8,22], as shown for example in Fig. 10(c). We can only speculate that maybe not all the modes of the input multimode core were excited in these experiments, and that there were fewer single-mode fibres between the lanterns than required by Eq. (8). If the ratios of mode numbers involved was appropriate (eg, $N_{SM}$ being intermediate between $N_{MM}$ and the number of modes excited at the input), a wide low-loss wavelength range could be achieved. If so then this was inadvertently the "light bucket" method at work, Fig. 20, but with the reduced input $N$ due to restricted excitation rather than waveguide design.

In any case, the wavelength dependence in Eq. (11) can be avoided by relaxing the requirement that the single-mode cores are identical and actually single-mode. The number $N_{MM}$ of modes in the multimode core increases with decreasing wavelength because additional higher-order modes become guided; this increase can be matched in the array of "single-mode" cores by making them multimode at appropriate wavelengths. The right profile of second-mode cutoffs within the population of "single-mode" cores can ensure mode-number matching over a considerable wavelength range. This idea is discussed in more detail in Section 4.2. Its drawback is that it can't be used to make MMF devices with SMF performance (such as FBG filters, Section 4.1), which require MM-SM-MM lantern pairs with intermediate cores that are identical and single-mode.

Another way to avoid the wavelength dependence of loss due to mode-number mismatch is to choose a multimode core design with a wavelength-independent number of modes. Such behaviour is seen in photonic crystal fibres, some designs of which are endlessly single-mode [135]. PCFs with holes that are slightly too big to be endlessly single-mode are endlessly no-more-than-two-moded [136], and this behaviour generalises to higher mode numbers for other PCF designs. The type #1 photonic lantern may therefore exhibit a wavelength-insensitive loss with identical single-mode cores, though this was not tested in [3]. However, most multimode systems that one might wish to connect it to will support a wavelength-dependent number of modes, in which case the loss will simply occur at the interface instead.

### 3.2. Symmetry and mode-order changes

Mode-number mismatch, as discussed in the previous section, is the most basic loss mechanism specific to photonic lanterns. Mode-number matching is necessary if an unpredictably- or incoherently-excited input set of modes is to map reversibly onto an output set of modes. However, it is not sufficient.

To say that $N$ modes are guided means that other modes are not guided: they are cutoff and become radiation modes. All the modes can be listed in decreasing order of their propagation constant $\beta$. We will assume that guidance is by total internal reflection, rather than by a photonic bandgap for example, which is true for all photonic lanterns that we are aware of. In that case the $N$ lowest-order modes with $\beta > kn_2$ (where $n_2$ is the cladding index and $k = 2\pi/\lambda$) are guided and the modes with $\beta < kn_2$ are not.

In structures with symmetry it is possible for modes of different symmetries to swap order along a transition. The $N$-th mode at the input may become the $(N+1)$-th mode at the output, and vice versa. In that case, even if the number $N$ of guided modes is matched along the transition, the light in the $N$-th input mode will be lost [2,36].

### 3.2.1. Example: square array

A toy model can be used to analyse the effect of mode-order changes analytically. Consider an $M \times M$ square array of weakly-coupled single-mode cores ($N = M^2$) that tapers into a strongly-confining multimode square core of refractive index $n_1$ and side $b$, Fig. 22. Assuming ideal $\psi = 0$ boundaries, the modes $\psi_{p,q}$ of the multimode square core are

$$\psi_{p,q}(x,y) = \sin\left(\frac{p\pi x}{b}\right)\sin\left(\frac{q\pi y}{b}\right) \tag{12}$$

where $p$ and $q$ are integers between 1 and $M$, and the origin of cartesian coordinates $x$, $y$ is at a corner of the square [137]. The supermodes $\psi'_{p,q}$ of the square array of single-mode cores [138] have amplitude and phase distributions across the cores that match a sampled version of $\psi_{p,q}$. Along a gradual transition that respects symmetry but is otherwise adiabatic, we therefore expect mode $\psi_{p,q}$ and supermode $\psi'_{p,q}$ to evolve into each other. The propagation constants in each case are

$$\beta_{p,q}^2 = k^2 n_1^2 - \frac{\pi^2}{b^2}\left(p^2 + q^2\right) \tag{13}$$

for the multimode square core and

$$\beta'_{p,q} = \beta_0 + C\left[\cos\left(\frac{p\pi}{M+1}\right) + \cos\left(\frac{q\pi}{M+1}\right)\right] \tag{14}$$

for the square array of single-mode cores, where $\beta_0$ is the propagation constant of an isolated core and $C$ is the nearest-neighbour coupling coefficient. To keep the model analytic (it is a toy model after all), we will simply assume that only the first $N = M^2$ modes are guided in the multimode square core so that the system is mode-number matched.

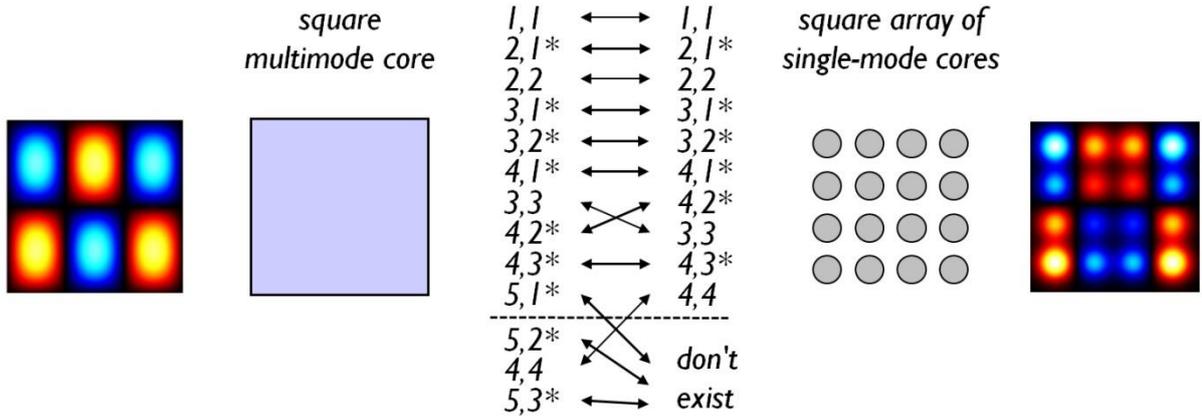

Fig. 22. (centre left) A strongly-confining multimode square core and (centre right) a 4 × 4 square array of weakly-coupled single-mode cores. (far left and far right) The (p, q) = (3, 2) mode of both, by way of example, where red and blue colours represent opposite phases. (middle) The (p, q) values of the lowest-order modes of both square systems according to Eqs. (13) and (14), listed in decreasing order of β. The asterisks identify degenerate pairs, and the dashed line marks where 16 modes are guided. The 16th mode of each system is not guided by the other system.

The differing functional forms of Eqs. (13) and (14) allows mode orders to swap. For example, (p, q) for the lowest-order modes of both systems are listed in decreasing order of $\beta$ in Fig. 22 for $M = 4$. The 16th mode of the multimode square core is (p, q) = (5, 1), which is not guided by the square array of single-mode cores, and the 16th mode of the square array is (p, q) = (4, 4), which is not guided by the multimode core. Even though the structure is mode-number matched, 1/16 of the light would still be lost.

### 3.2.2. Symmetry loss in the limit of large N

The toy model can be taken further to find an estimate for the minimum loss of a photonic lantern in the large-$N$ statistical limit. Much like phase-space arguments for calculating densities of states in solid-state physics [139], the modes in the model lantern can be mapped on a transverse $k$-space proportional to an integer grid of (p, q) values. The supermodes of the square array of single-mode cores occupy a square in (p, q) space, since p and q each take integer values from 1 to M. If we allow p and q to take negative values as well (thus quadruple-counting) and also accept zero values (whose relative contribution tends to zero in the limit of large N), then the supermodes occupy a square Brillouin zone in the $k$-space. The modes of the multimode square core occupy a circular quadrant with a radius proportional to a V-value like $V$ in Eq. (3). This follows from imposing $\beta > kn_2$, the condition for total internal reflection, onto Eq. (13). Again quadruple-counting, this becomes a circular disc in the $k$-space.

The symmetry loss arises because the square and the disc will never overlap perfectly, Fig. 23(a). The circular zone can be made larger or smaller by adjusting $V$ (via the core size or wavelength, for example), but there will always be some area of the square that is outside the disc, or some area of the disc that is outside the square, or both. Since these areas are proportional to numbers of modes, simple geometry combined with the reasoning leading to Eq. (9) can be used to plot expected loss versus the relative sizes of the circle and the square. Where one shape lies entirely within the other, the calculated loss matches Eq. (10), but in between the loss is greater than predicted by Eq. (10).

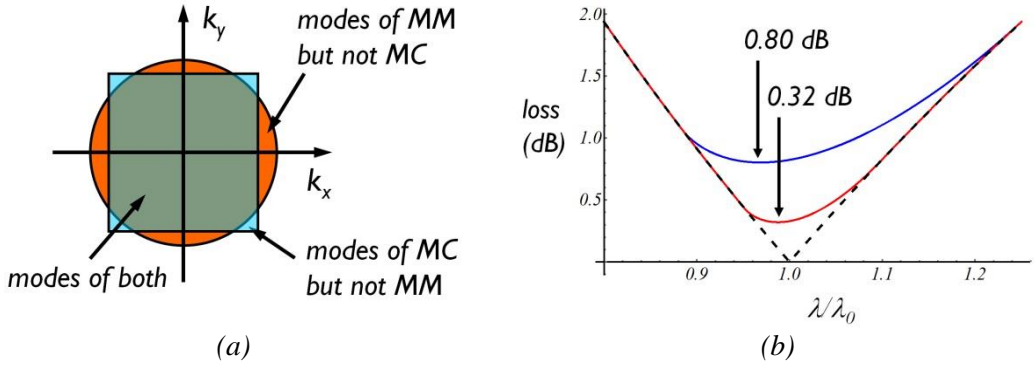

*(a)* *(b)*

*Fig. 23. (a) A circular disc representing the modes of the multimode square core and a square representing the supermodes of a square array of single-mode cores, plotted in k-space. The size of the circle can vary (eg with wavelength λ) but there are always parts of one shape that don't overlap the other. (b) Loss versus $\lambda/\lambda_0$ for a square array (blue line), a triangular array (red line) and ignoring symmetry loss (dashed line) [2]. The second-mode cutoff wavelength of the SM cores is assumed to lie outside the wavelength range.*

The resulting loss is plotted in Fig. 23(b) for a reversible MM-SM-MM lantern pair, together with Eq. (11) assuming no symmetry loss. The horizontal axis is a normalised wavelength $\lambda/\lambda_0$, where $\lambda_0$ is the wavelength where Eq. (11) gives zero loss, and the wavelength dependence of $V$ is given by Eq. (3). The minimum loss is now 0.80 dB rather than 0 dB (as plotted in Fig. 21), though it still lies near $\lambda_0$.

The exercise can be repeated but for the single-mode cores in a triangular array. This gives the Brillouin zone a hexagonal shape. Since a hexagon is more circular than a square is, there is a smaller range where one shape doesn't enclose the other, and a lower minimum loss of 0.32 dB [2].

A general approach to minimising the symmetry loss may therefore be to adopt a quasi-crystalline or golden-angle spiral array of cores, which would have a Brillouin zone that is more circular than a hexagon is [140,141]. However, since the problem is caused by symmetry, a simpler solution would be just to avoid symmetric patterns of cores. Recall that the loss arises because some modes change order along the transition: in the absence of symmetry, modes can't change order and anti-crossings form instead. However, the anti-crossings may be so narrow that a practical transition is not gradual enough to be adiabatic with respect to mode coupling across the anti-crossing.

*3.2.3. Eliminating symmetry loss*

Fontaine et al [36] proposed a method for eliminating symmetry loss completely. The cores in the single-mode array are positioned in such a way as to allow the supermodes to more-closely resemble sampled versions of the modes of the multimode core. These do not match the periodic patterns of cores that naturally arise by the stacking of identical fibres (or, in the case of multicore fibres, identical canes used to assemble the fibre preform).

Specific examples were given of core patterns, together with simulations demonstrating the contrast between well-chosen and poorly-chosen geometries. For example, Fig. 24 shows the lowest-order modes of a circular core, together with the supermodes of two different patterns of $N = 15$ cores. In one case, each supermode has a guided partner among the 15 lowest-order modes of the multimode core, so there is no change of mode order along the transition and no symmetry loss. In the other case, the 15th mode of each system is not guided by the other, which would lead to a loss of 0.3 dB.

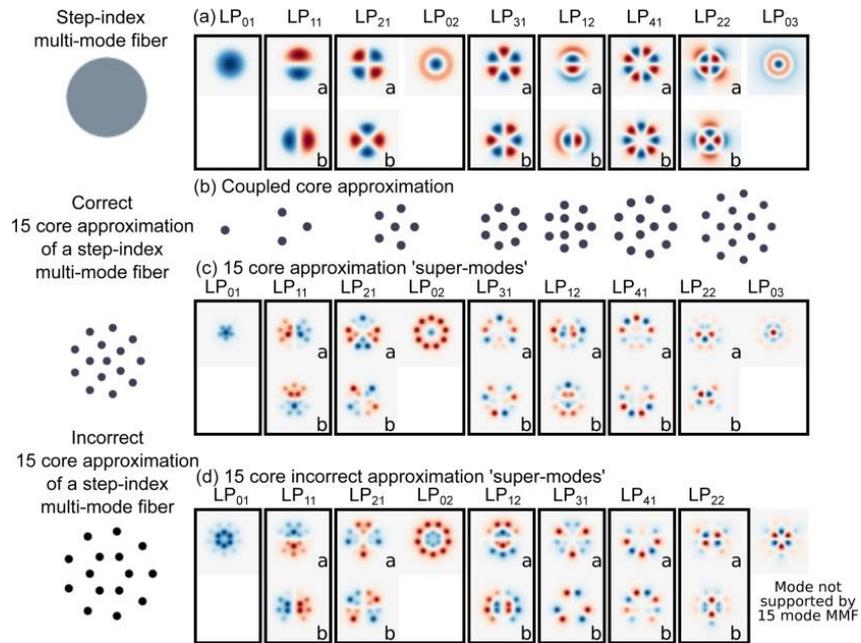

*Fig. 24. (a) Calculated mode patterns of the 15 lowest-order modes of a circular fibre core. (b) Core patterns that match successively-bigger sets of those modes. (c) Supermodes of a well-chosen array of single-mode cores. (d) Supermodes of a poorly-chosen array of single-mode cores. Reprinted from [36].*

The avoidance of crossings between guided and unguided modes is illustrated in Fig. 25 for the larger value of $N = 51$. In the well-chosen pattern of cores there is no swapping between guided and unguided modes, as can be seen by the clear gap between the two sets of curves. In the poorly-chosen pattern of cores, two modes swap over.

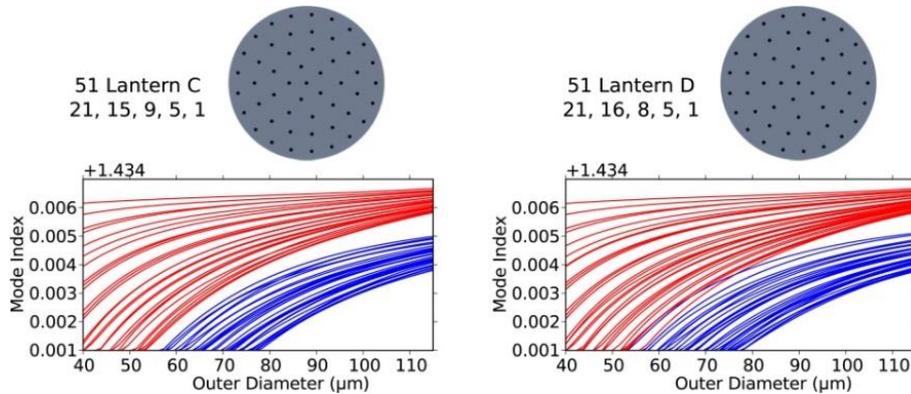

*Fig. 25. Calculated variation of mode indices versus cladding diameter for two tapered lanterns with $N = 51$ cores. Red curves are guided supermodes whereas blue curves are cladding modes. The pattern of cores is well-chosen for lantern C and poorly-chosen for lantern D. Reprinted from [36].*

Despite the attention we have just given to symmetry loss, we should emphasise that it is rather small. The loss of one mode out of 51 is less than 0.1 dB according to Eq. (10). Furthermore, the highest-order guided mode of a multimode core is close to cutoff and therefore likely to be lost anyway for reasons unconnected with the lantern, such as bend loss. However, for space-division multiplexing in telecoms the complete loss of one mode has an effect that is out of proportion to the amount of light it carried, including the possible loss of a complete channel of information [36].

*3.3. Adiabaticity of the transition*

Mode-number mismatch and symmetry-induced swapping of mode order, as discussed in Sections 3.1 and 3.2, cause loss even in a lantern transition that is perfectly gradual. In practice, mode-coupling to higher-order modes takes place if the transition is not gradual enough, as discussed for a single core in Section 2.1.2. Since this light leaves the set of $N$ modes that are guided in the output system, it is lost.

Various theoretical criteria exist to place conditions on the shape of a taper transition if it is to be adiabatic [142,143]. Which criterion to use depends on the desired balance between simplicity and accuracy. However, what they all agree on is that, for a given disturbance, coupling is more likely between modes that are close in $\beta$. For example, the "weak power transfer" criterion [25,52,143] can be expressed as

$$\left| \frac{2\pi}{(\beta_1 - \beta_2)} \frac{d\rho}{dz} \int \Psi_1 \frac{\partial \Psi_2}{\partial \rho} dA \right| \ll 1 \tag{15}$$

where $\Psi_1$ and $\Psi_2$ are the normalised field distributions of the local modes between which power coupling is most likely, $\beta_1$ and $\beta_2$ are their respective propagation constants, $\rho$ is a parameter (such as a local core radius) representing the transverse size of the local waveguide and $z$ is the co-ordinate along the fibre. For a given physical transition $d\rho/dz$ and similar values for the integral (the most important role of which is to rule out coupling that is forbidden by symmetry), the criterion is hardest to satisfy when $\beta_1$ and $\beta_2$ are most similar.

In the array of single-mode cores, the supermodes all have propagation constants close to that of an isolated SM core and far from those of radiation modes. Guided-mode and unguided-mode $\beta$'s become closer together at the tapered-down multimode end of the structure. This behaviour can be seen in Fig. 25, for example. On the other hand, once the light has filled the MM core and the residual SM cores can be neglected, reducing the size of the MM core increases the $\beta$ separation. This can for example be seen in Eq. (13) for a strongly-confining square core as $b$ decreases, behaviour that is entirely typical of all MM waveguides [144]. The part of the transition that is most sensitive to mode-coupling loss is therefore the point where the light has spread out from the SM cores but not yet become well-guided by the MM core [94].

An analytic account of how loss scales with the number of cores $N$ can be obtained using the toy model of Section 3.2 and the result of the previous paragraph that mode-coupling loss is most likely where the light has left the SM cores. We consider coupling from the $N$ lowest-order modes of a strongly-confining MM square core to the $(N + 1)$-th mode: the first that isn't guided as a supermode in the multicore array. In the spirit of Eq. (15), light in modes whose $\beta$ values are closer than a threshold $\Delta\beta$ (given in principle by the geometry of the waveguide and its axial rate of change $d\rho/dz$) is assumed to couple completely out of the guided set, while light in the other guided modes remains where it is [2]. To avoid speculating about the threshold, dividing the estimated loss by that for the case of $N = 2$ normalises $\Delta\beta$ (and indeed the profile of the taper transition) out of the problem. $N = 2$ is a useful benchmark because (as discussed in Section 2.1.2) it corresponds to the 2 × 2 fused coupler, a widely-available fibre component that can readily be made with low loss for taper transitions a few millimetres long [81-87]. Increasing the length of the transition decreases the threshold $\Delta\beta$ and allows more modes to avoid loss. We increase $N$ in proportion to the square core's area $b^2$, Eq. (13), so that the "unit cell" occupied by each core is unchanged in size and the integrals in Eq. (15) are similar. This packs more modes into the range $\Delta\beta$, but reduces the fraction of the total power in each. Thus the fraction of the total that experiences loss is unchanged, suggesting that for large-enough $N$ the relation between loss and transition length approaches a universal function. The result of such a calculation is shown in Fig. 26 for $N = 20$ and $N = 500$. We see that a transition several times longer than that in a 2 × 2 fused coupler may be needed, but $N$ can otherwise be increased indefinitely without penalty [2].

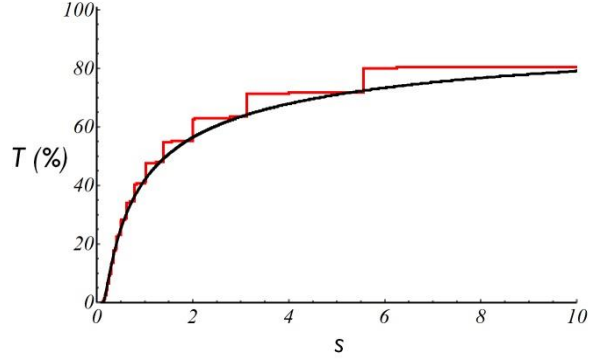

*Fig. 26. Estimated transmission T along a transition with N = 20 (red) and N = 500 (black) notional cores, versus the transition length s normalised to the length that gives no loss for N = 2 [2]. The cores are taken to be the same for all three values of N, and cladding area $\propto N$.*

We should also consider how to improve losses for a given *N*. We found that decreasing the SMF's cladding-to-core ratio (eg using fibres with reduced overall diameters) relaxes the requirements for adiabaticity and reduces loss for a given *N* [52], Fig. 11. Huang et al have reported that replacing the SMFs with graded-index MMFs, in which only the fundamental modes are excited, also promotes adiabaticity [76]. Presumably graded-index MMFs of reduced diameter would be better still.

For quantitative predictions of mode-coupling loss it is necessary to resort to a numerical simulation. The beam propagation method (BPM) is particularly suitable for such calculations because it is designed for use on non-uniform waveguide structures like taper transitions. Fig. 27 is a BPM simulation of the loss through an MM-SM-MM lantern pair (shown in Fig. 4) made by tapering a multicore fibre with $N_{SM} = 85$ cores separated by a pitch of $\Lambda = 17$ μm [25]. The input field was made up of all the guided modes with equal amplitudes and random phases. The loss was calculated for different values of the transition length *L*, by overlapping the calculated output field with the guided modes and assuming that light not in any of the mdoes is lost. The loss is high when the transition is short, as expected, but reduces to below 0.35 dB if the transition length is 2 cm or more. 0.25 dB of this loss can be accounted for by a mode-number mismatch via Eq. (10) because the multimode core in the model actually supported $N_{MM} = 90$ modes rather than 85.

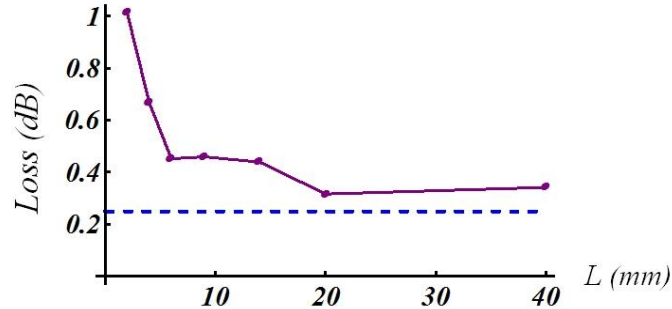

*Fig. 27. Simulated dependence of loss on transition length L for the multicore lantern pair of Fig. 4. The broken blue line represents the 0.25 dB of loss that is due to a known mismatch in mode number. Reprinted from [25].*

Thus realistically-long transitions in a MCF with closely-spaced cores can yield very low loss, in contradiction to [9] which concluded that the SM cores must be at least $\Lambda = 60$ μm apart. This is because the cladding diameter in [9] was fixed at 110 μm as $\Lambda$ varied, so that most of the fibre was empty of cores when $\Lambda$ was small. When such a fibre is tapered, the field expands from a small central array of cores to fill the entire fibre, a dramatic change causing a high value for the integral in Eq. (15) and hence a high loss. In contrast, the cores in the MCF of Figs. 4 and 27 are distributed throughout the fibre, so that the field in each core only needs to expand enough to occupy its own unit cell in the cross-section, a much more gradual change of field shape. Thus the high loss reported in [9] was due to an unrepresentative feature in the modelled MCF.

## 4. Applications of photonic lanterns

The purpose of a photonic lantern is to provide a low-loss interface between a multimode waveguide and several single-mode waveguides. It is best suited for situations where control over the distribution of light between the output degrees of freedom (modes of a multimode core, or cores in an array of single-mode cores) is unimportant. Photonic lanterns therefore have several applications where such an interface is needed.

### 4.1. Single-mode performance in multimode fibre

One broad class of applications is where a type of optical waveguide device has high performance in single-mode fibre (SMF) but doesn't exist, or performs poorly, for multimode fibre (MMF). This can arise because the various modes of an MMF have different propagation constants, different spatial distributions and different susceptibilities to sources of loss such as bending loss. If the light can be transferred into several identical SM cores, such mode dependence is eliminated because all of the light now travels in an identical way.

A good example of such a waveguide device is the fibre Bragg grating (FBG). A periodic variation in refractive index can be written along an SMF core by exposure to an interferometrically-generated pattern of ultraviolet light [145]. If the propagation constant $\beta$ of the light and the period $\Lambda$ of the variation satisfy a Bragg condition

$$\beta = \frac{\pi}{\Lambda} \qquad (16)$$

then light reflected at each period interferes constructively to produce a large reflected signal. Since $\beta$ depends on the wavelength of the light while $\Lambda$ is a constant, the Bragg condition is satisfied around a single Bragg wavelength, and other wavelengths are transmitted without reflection. The FBG therefore functions as a simple but effective spectral filter, strongly reflecting a well-defined narrow band of wavelengths while transmitting the rest. Such single-mode FBGs are widely manufactured, for use in fibre lasers and sensors for example. More complicated spectral responses can be achieved by writing an appropriately aperiodic index variation along the core [103]. Each Fourier component of the index variation then causes a particular wavelength to be reflected.

It is of course possible to write an FBG in multimode fibre. The Bragg condition Eq. (16) then applies separately for each guided mode, with its own value of $\beta$. The wavelength at which light is reflected therefore depends on the mode it propagates in. If the light is carried in many modes (as is usually the case), reflection will be weak, distributed across a number of wavelengths, and dependent on the mode content of the light [104]. Applications for such a response are likely to be highly specialised, if not contrived.

An MM-SM-MM lantern pair provides an effective way to impose the strong well-defined spectrum of a single-mode FBG on a multimode fibre, Fig. 5. The entire structure is an optical device with MMF ports that can be connected readily to MMFs. However, the light passing through the device encounters the FBGs in SM cores. The light in all the cores is acted upon by the same spectral response, so the reflection and transmission spectra of the device as a whole match those of the individual single-mode FBGs. The first photonic lantern experiment [1,3] demonstrated this idea for 19 identical simple (ie, periodic) FBGs each spliced between SMF ports of two of the type #1 lanterns described in Section 2.2.1. The spectral transmission between the two MMF ports is shown in Fig. 7(a), which looks very like a single FBG response.

The motivation for these experiments was the need in astronomy to pre-filter infrared light collected by telescopes. Hydroxyl (OH) groups in the upper atmosphere store energy from sunlight during the day and radiate it as IR light during the night. This light swamps ground-based observations of the faint highly-redshifted light from the early Universe, for example. Since this OH emission is concentrated in a number of very narrow intense spectral lines, it can be filtered out by suitable aperiodic FBGs in SMFs [103]. However, the light collected even from a point source by a seeing-limited telescope is not diffraction-limited (in the absence of adaptive optics) and so couples poorly to SMFs. A suitably-designed photonic lantern with aperiodic FBGs in the SMFs can effectively filter out the unwanted emission lines from an MMF carrying light from the telescope, allowing observations to be made using the portions of the spectrum in between, Fig. 28. The PIMMS schemes combine such a lantern filter with the formation of a pseudo-slit (Section 4.3) input to a spectrograph or an arrayed waveguide grating to produce powerful but compact and economical spectroscopic instrumentation [10,23,49,61], and these ideas are being tested on-sky [21,26,28,31,34,39,66]. This is an important application in the field of astrophotonics, which seeks to take advantage of photonics technology (as developed for telecommunication for example) to solve problems in observational astronomy [23,146].

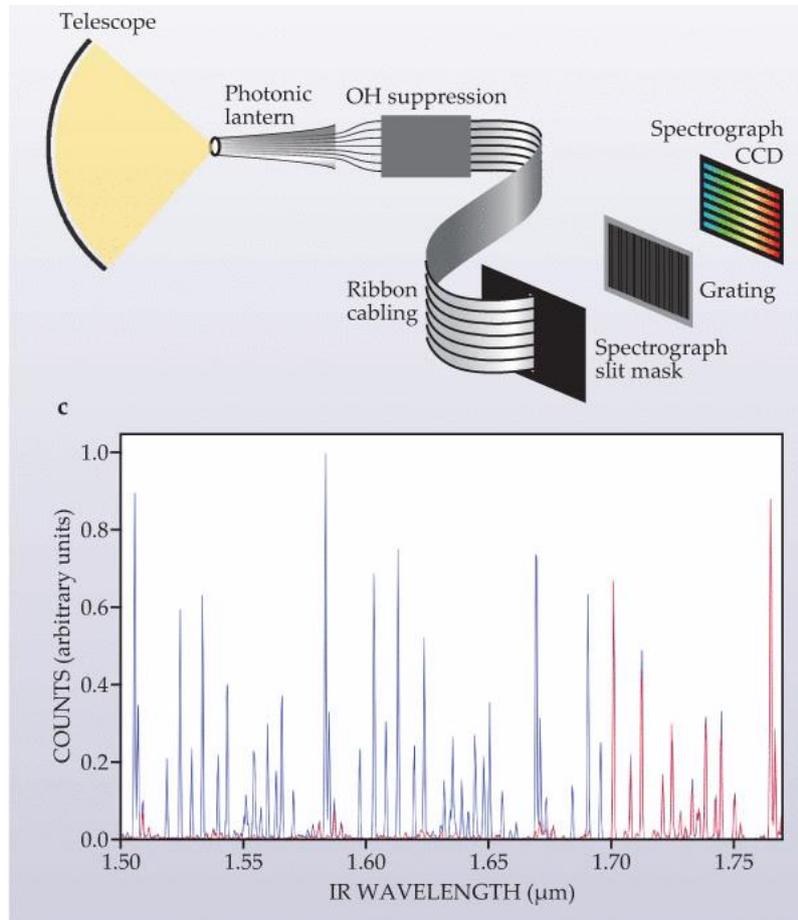

*Fig. 28. (upper) PIMMS #0, where the output SMFs of a multi-fibre lantern are positioned in a row to form a pseudo-slit at the entrance to a spectrograph. (lower) Spectrum of OH emission lines before (blue) and after (red) removal below 1.7 μm wavelength by a lantern-based FBG filter. Reprinted with permission from J. Bland-Hawthorn et al Phys. Today **65**(5), 31 (2012) [23]. Copyright 2012, S. Ellis.*

For reasons outlined at the beginning of Section 2.2.3, it would be desirable to make multimode FBG filters from multi-core type #3 lanterns rather than multi-fibre type #2 lanterns. The manufacturability and scalability advantages discussed there would be further enhanced by the prospect of writing the FBGs across all the $N$ cores of the multicore fibre at once, rather than $N$ times in $N$ separate SMFs. In [12,18,25] we reported low-loss multicore fibre lanterns, and found that FBGs could be written across most or all of the MCF's 120 cores by the same technique as normally used to make FBGs in single-core SMFs. However, we had limited success in demonstrating the required uniformity of FBG transmission spectra, so that the complete MMF filter's response did not match the best individual SMF spectra, Fig. 29. Work is currently underway to solve these problems, and so demonstrate the type #3 MCF lanterns as the most scalable and manufacturable approach to make MMF FBG filters particularly for astronomy [66,147-150]. The combination of Bragg-gratings with type #4 ULI lanterns has also been proposed [4] and recently demonstrated [54].

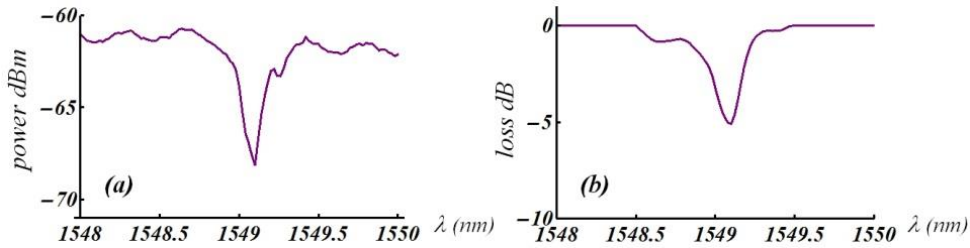

*Fig. 29. (a) Output spectrum from a type #4 MM-SM-MM lantern pair made from the N = 120 MCF of Fig. 13, with a simple FBG written across all of the cores. (b) The mean of the notch responses of the 120 individual cores. Reprinted from [25].*

Another example of using photonic lanterns to achieve single-mode performance in a multimode device is the mode-selective switch of [55], where lanterns allow a common wavelength-selective switching response for the light in all modes. Photonic lanterns can be used to assist detection by splitting multimode light from free space between several SMFs suitable for coherent detection [44] or, conversely, by merging signals from separate SMFs onto a single detector [16]. Lantern-like devices have also been proposed as brightness-conserving incoherent [6,14,15] or coherent [151] combiners for laser beams.

*4.2. Multicore fibres as multimode fibres*

Not only are most SMF components superior to their MMF counterparts, but in most respects the fibre itself is superior. The original motivation for SMF, in a world familiar with MMF, was the elimination of intermodal dispersion. The different modes of the MMF have different group velocities (or differential group delay, DGD), so a pulse of light at the input of the MMF will be spread in time when it reaches the output via the different modes. Graded index fibres reduce the problem [152], but (aside from the much-smaller polarisation mode dispersion) it is of course entirely eliminated by having just one mode in SMF. DGD also increases the complexity and expense of MIMO signal processing in space division multiplexing, discussed in more detail in Section 4.4.

Another key disadvantage of MMFs is bend loss. Bending causes some light in a fibre mode to be radiated into the cladding. Since the mode spectrum of an MMF fills the effective index range between the core and cladding refractive indices, the highest-order modes (those closest to cutoff) will be very close in index to the radiation modes of the cladding. Not only will they be highly susceptible to bend loss [153] (recall from Section 3.3 and Eq. (15) that coupling becomes more likely between modes of similar effective index or $\beta$), but they also provide "stepping stones" via which light is coupled into the cladding from modes further from cutoff, Fig. 30. Thus as an equilibrium mode distribution is set up [152], light will continually leak out of the fibre.

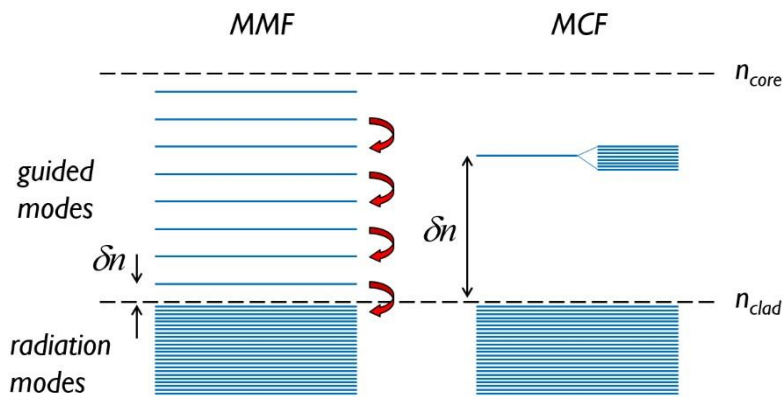

*Fig. 30. (left) Schematic mode spectrum of a multimode fibre (MMF), filling the range between core and cladding refractive indices. Light in the modes closest to cutoff leaks out of the core due to bend loss because the effective index step $\delta n$ is so small, to be replaced (in an equilibrium mode distribution) by light from modes further from cutoff. (right) Schematic mode spectrum of a multicore fibre (MCF), comprising a large set of near-degenerate modes around the effective index of a single-mode core (or a few dissimilar ones), which can be far from the cladding index.*

In astronomical applications, the redistribution of light among the modes causes a problem called focal ratio degradation, FRD [134]. If light from telescope optics with a certain focal ratio $f/\#$ under-fills an MMF (that is, the fibre's *NA* exceeds the angularly-matched value given by $f/\# = 1 / 2\,NA$), then light will be coupled along the MMF until it fills its mode spectrum. The output wave therefore has a smaller (ie, photographically "faster") focal ratio which demands more expensive free-space optics to handle, and ultimately a more expensive spectrograph. If instead the input light is coupled into a lower-NA MMF whose mode spectrum it fills, then it will suffer greater bend loss for the reason described in the previous paragraph. Clearly losses are undesirable in astronomy, where "every photon counts". The problem is exacerbated by the large focal ratios typical of astronomical instrumentation, meaning that the MMF would need an unusually-small NA and so would suffer worse bend loss in any case.

There is still however a place for MMFs because they can carry so much more light than SMFs, in terms of power or number of modes. They are easier to use, and are well-suited to multi-mode light sources (such as seeing-limited telescopes) that are not diffraction-limited. We thus come to the point that originally motivated the photonic lantern - to exploit the advantages of SMFs but for multimode light - and so can consider whether type #3 photonic lanterns would allow us to use multicore fibres as multimode transmission fibres but without the intermodal dispersion and high bend loss. Lanterns allow the input and output of an MCF to interface to ordinary MMF systems, but the light is transmitted along single-mode cores. Assuming that the lantern's MMF input port is mode-number matched to the MCF, input light filling the NA will be transmitted to the output with no DGD (other than from differences between the cores due to bends and manufacturing imperfections) and a flat mode spectrum. It will also suffer from no more bend loss than an SMF over all the guided modes, Fig. 30, and no FRD.

One apparent disadvantage of this scheme is that the number of single-mode cores in the MCF is fixed whereas the number of modes guided by an MMF depends quite strongly with wavelength, as discussed in Section 3.1.2. The mode-number matching condition Eq. (8) is therefore only satisfied at a single wavelength, with loss rising sharply either side, Eq. (11) and Fig. 21. However, this discrepancy in *N* can be compensated by designing the cores of the MCF so that some become multimode in the wavelength range of interest [60]. A set of cores with the right diameter distribution will guide the right number of extra modes between them to keep pace with the increasing number of modes guided by the MMF.

To calculate the diameter distribution we recall the wavelength dependence of Eq. (6), and note that the so-called "second mode" is in fact a pair of degenerate $LP_{11}$ spatial modes (not counting polarisation states separately). The number of modes in a core therefore rises from 1 to 3 (not 2) below its second-mode cutoff wavelength $\lambda_{co}$. Furthermore at the wavelength $0.63\,\lambda_{co}$ a step-index core will start to guide the "third mode" group (two $LP_{21}$ modes and one $LP_{02}$ mode) as well, raising the number of modes from 3 to 6.

The results of such a calculation are presented in Fig. 31 for an MCF with $N = 73$ cores, though the results are quite general aside from discretisation effects like steps in the loss spectrum. We assume that all *N* cores are single-mode at $\lambda_L$, the longest wavelength of interest, and so can match the MMF port of a lantern with *N* modes. To match the number of MMF modes at shorter wavelengths, the cores should have the distribution of diameters (and hence second-mode cutoff wavelengths) plotted in Fig. 31(a), the diameter of core 73 (the smallest) being about half that of core 1 (the biggest). There is a kink at core number 57 because it starts to guide the second mode as core 1 starts to guide the third mode. The total number of modes in the array of 73 cores and in the matching MMF port of the lantern are plotted in Fig. 31(b), and Fig. 31(c) shows the expected loss as calculated using Eq. (11). We see that close mode-number matching and hence low loss can be sustained over a spectral octave, from $\lambda_L$ down to about ½ $\lambda_L$. (A wider range is possible if there are fewer cores with more modes each, though this eventually leads to the uninteresting limit of a single core with all the modes: the MMF we originally sought to replace.)

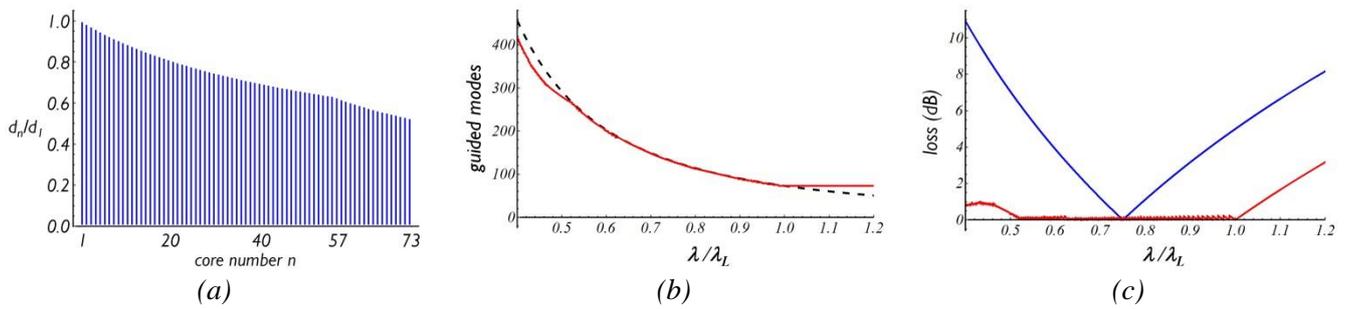

*Fig. 31. (a) Distribution of core diameters $d_n$ in an $N = 73$ MCF, such that the total number of guided modes matches that of a MMF over a broad wavelength range. (b) Wavelength dependence of the number of modes guided by the MCF (solid line) and its matching MMF (broken line), where $\lambda_L$ is the second-mode cutoff wavelength of core 1. (c) Loss due to mode-number mismatch of an MM-SM-MM lantern pair made using the MCF (red), together with the corresponding spectrum for an MCF with identical cores optimised for $\lambda = 0.75\, \lambda_L$ (blue). For the blue curve, the second-mode cutoff wavelength is assumed to lie outside the wavelength range.*

Fig. 32 shows images of an $N = 511$ MCF made to exploit this principle and the MMF port of a photonic lantern made from it [60]. The fibre was made by first assembling 7 canes each containing 73 cores with the diameter profile of Fig. 31(a), then stacking the canes to form the preform for the final MCF. The fibre was designed to make photonic lanterns to match an MMF with a core diameter of 55 µm and NA of 0.23 between the wavelengths of 380 nm and 860 nm. It was designed to act as a multimode relay fibre between a telescope and a spectrograph with low loss and low FRD, but was also intended to act as a mode scrambler.

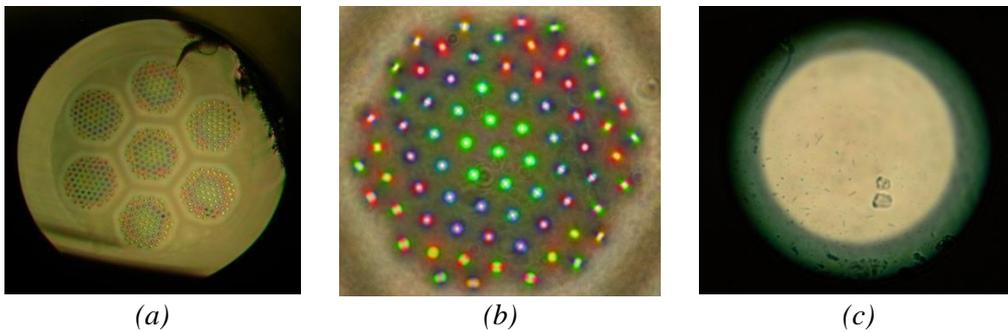

*Fig 32. (a) An experimental MCF with 511 cores, in seven clusters of 73 cores with the diameter distribution of Fig. 31(a). The fibre diameter is 600 µm. (b) A single cluster of 73 cores illuminated by white light, the different colours and mode patterns in the cores being indicative of their varying diameters. (c) The multimode port of a photonic lantern made by tapering the fibre in an F-doped jacket to a diameter of 55 µm. The images are not to the same scale.*

At the output of an ordinary MMF excited by a coherent source, the intensity distribution depends both on how the fibre's modes are excited at the input and also on the phase differences they acquire as they propagate. Such a variable output pattern can cause modal noise if the response of the instrument at the output (eg, a spectrograph) depends on it. To make the output incoherent it is necessary either to make the fibre so long that the modal phases are spectrally averaged over the bandwidth of the light, or to continuously disturb the fibre to time-average the phases. Even so, this only changes the phases of the modes and not their amplitudes. In contrast, effectively-random phase changes due to bends in the MCF couple light between the supermodes and hence the modes of the output MMF, scrambling the modal amplitudes as well as their phases [25], as depicted in Fig. 18. This is helpfully exaggerated by any dissimilarities between the cores, for example those introduced deliberately in the design of the fibre of Figs. 31 and 32. This encourages the development of an incoherent and uniform mode spectrum, with the aim of producing an output pattern that is insensitive to the light pattern at the input.

Experiments with the $N = 511$ dissimilar-core MCF are underway [60], but mode scrambling was previously investigated with the $N = 120$ identical-core MCF discussed in Section 4.1 and presented in Fig. 33. The output of a graded-index MMF carrying 1550 nm light from a diode laser was used as the light source. This was butt-coupled to ~2 m of simple step-index MMF and, alternatively, to a similar length of an $N = 120$ MCF with type #3 photonic lanterns at both ends [25]. The MMF ports of both lanterns matched the core diameter and NA of the simple MMF. The outputs imaged in the near field using an InGaAs camera are shown in Fig. 33(b) and (c), while Fig. 33(d) shows the output of the lantern-terminated MCF while a ~1 m loop of the MCF was gently oscillated. The pattern at the lantern's output had the high spatial frequencies expected for modes close to cutoff in a core guiding 120 modes, whereas the lower spatial frequencies from the simple MMF matched those of the input field. This suggests that power in the MMF was not significantly coupled from the restricted set of modes initially excited, whereas the lantern distributed light across the full spectrum of guided modes. The time-averaged image shows how readily a modest perturbation can wash out all interference between the guided modes. Indeed the output illumination was so uniform and incoherent that the pattern of residual SM cores in the lantern's tapered-down MMF core can be discerned.

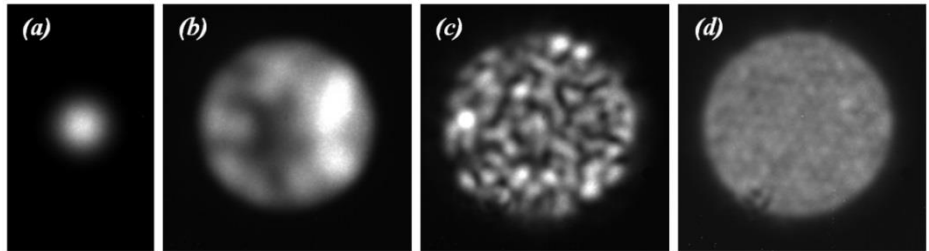

*Fig. 33. Near-field images at 1550 nm (time constant ~30 ms) at the output of (a) a graded-index MMF butt-coupled to (b) a step-index MMF or (c) an N = 120 MCF photonic lantern with similar step-index MMF ports. (d) repeats (c) while the fibre was gently disturbed and the camera time constant was increased to 0.1 s. Reprinted from [25].*

In contrast to multicore fibres, the paths of the individual SM cores in type #2 (multi-fibre) and type #4 (ULI) photonic lanterns can be defined almost arbitrarily. This means that the path lengths can be dramatically different within a single device, making mode scrambling very efficient. The effect of mode scrambling has been observed in MM-SM-MM lantern pairs made by both techniques [32,34,63,64].

*4.3. Reformatting multimode light*

Reformatting refers to changing the shape of the light pattern in a waveguide. For example, the image of a star at the output of a telescope is circular, at least in the sense of having an aspect ratio of 1:1. In contrast the ideal input to a spectrograph is slit-shaped, with a small width along the direction of the dispersion of its diffraction grating or prism. A wide slit (or fibre feed) compromises the resolution of the spectrograph, and converts modal noise in the fibre into apparent spectral noise in the measurement. In such circumstances it is advantageous to use a waveguide device to reformat a circular input to a high-aspect-ratio output. This is simple to achieve for single-mode light. An adiabatic transition between a 1:1 core and a slit-shaped core ensures that the aspect ratio of the guided mode is suitably transformed with low loss [88,90].

For multimode fibre supporting $N$ modes, the ideal case would be for the output light to have a diffraction-limited distribution in the $x$ direction across the slit [10,23,27,29,49,50,56,63]. This means the waveguide is single-mode in that direction, with the intensity distributions of all its modes having one lobe along $x$. All the field variations would instead be distributed in the $y$ direction along the slit, each mode having between 1 and $N$ lobes in that direction.

Unfortunately a simple waveguide transition (eg a taper) between these two limits would be very lossy. The problem is illustrated most clearly for an idealised rectangular geometry, Fig. 34. The $x$ and $y$ field variations of the modes are separable, and the input modes have between 1 and $\sqrt{N}$ lobes along both $x$ and $y$, Eq. (12). A gradual symmetric change in aspect ratio will maintain the shape of each mode but compress it along $x$ and stretch it along $y$, thus retaining the numbers of lobes in each direction. Since the output has only one lobe along $x$, all the modes with more lobes along $x$ must become cut-off at some point along the transition. In compensation, modes with more than $\sqrt{N}$ lobes along $y$ become guided, but these are never excited. The result is the loss of most of the light, with a surviving throughput of just $\sim 1/\sqrt{N}$. A realistic transition may of course not be of a rectangular waveguide, but the same problem of loss caused by the inhibited redistribution of degrees of freedom between the two axes remains.

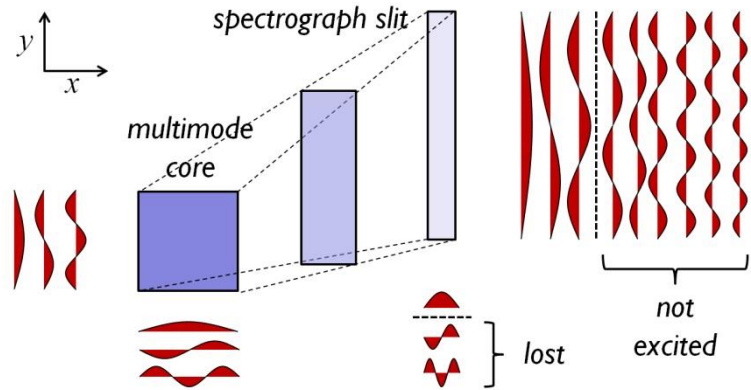

*Fig. 34. Idealised rectangular aspect-ratio transition between a 1:1 square core and a high aspect-ratio slit-shaped core, for N = 9. The 9 modes of the square core are products of the 3 x and 3 y field distributions shown, and try to evolve along the gradual transition into the corresponding modes of the slit-shaped core. However, that core only supports the products of the one-lobe distribution along x with the 9 y field distributions shown. Hence only 3 of these 9 modes are excited, and the other 6 input modes are lost.*

Photonic lanterns of types #2 or #4 can solve this problem, Fig. 35. In both cases the output SM cores can be positioned freely in three dimensions to form a pseudo-slit [134]. The SM cores in a ULI-inscribed type #4 lantern can be routed during the fabrication of the device [27,50,56,63], while the individual output fibres in a tapered multi-fibre type #2 (or indeed type #1) lantern can be positioned at will afterwards [10,23,49]. By positioning the cores in a row, a diffraction-limited output along *x* with low loss is guaranteed provided bending losses are avoided en route. In effect, the 3D routing of the cores creates a highly non-symmetric waveguide transition by almost-literally picking up the *x* degrees of freedom and distributing them along *y*.

The PIMMS #0 concept [10,23,49] exploits this idea in a multi-fibre type #2 lantern, in combination with OH-suppressing FBGs, Fig. 28(a). The device is intended to deliver seeing-limited light from a telescope to the entrance slit of a spectrograph. One disadvantage is that the output cores are inevitably separated along the *y* direction by their claddings. This problem can be reduced by using narrower SMFs which can be closer-packed, as shown in Fig. 11, and eliminated completely in ULI-inscribed type #4 lanterns by routing the cores so that they touch and merge in the final linear array [56,63], Fig. 35.

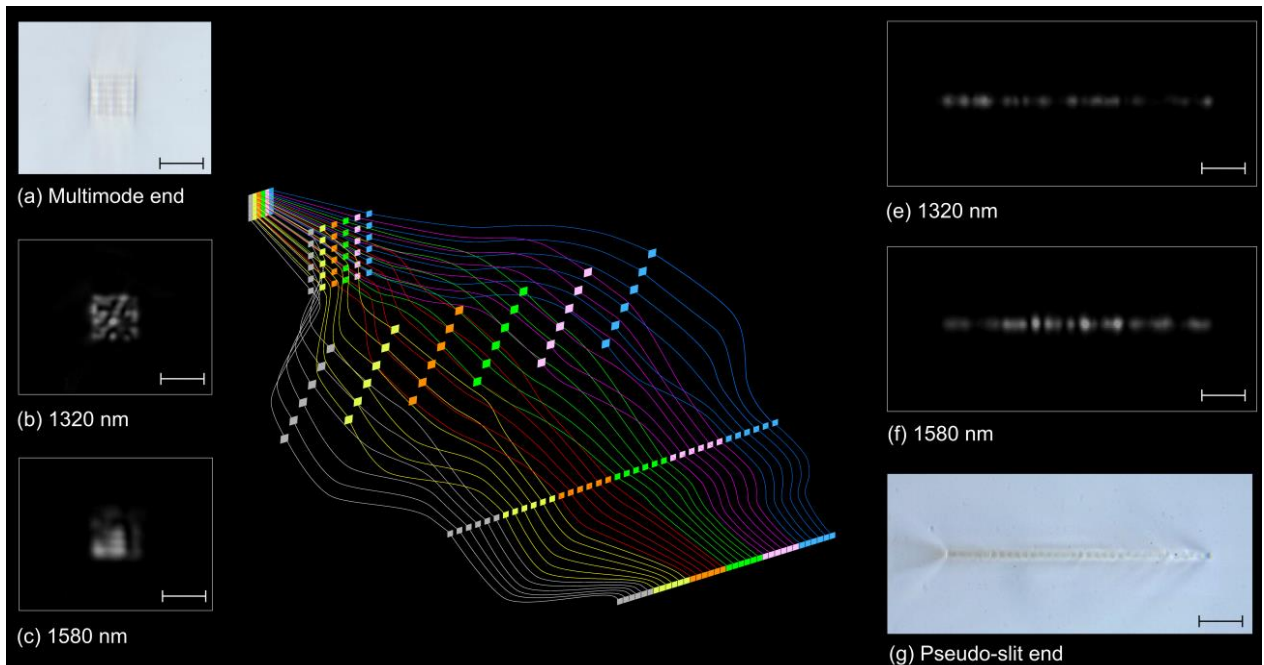

*Fig. 35. The "photonic dicer", where the output SM cores of a ULI-inscribed lantern are positioned in a row then merged to form a true slit-shaped output [56]. The insets are optical micrographs of the structures and light patterns at each end.*

Such reformatters cannot be made using type #3 lanterns based on tapered multicore fibres because the SM cores are held fixed in a 2-D array. However, a hybrid reformatting lantern can be made by combining a type #3 fibre lantern with a ULI-inscribed fan-out [13,24]. This takes advantage of the former's low loss and automatic connection with an MCF relay fibre and the latter's versatility in producing an arbitrary output, at the expense of the need for a low-loss junction between the two.

Another approach for making type #3 multicore lanterns work with the PIMMS #0 concept is the Photonic TIGER spectrograph [61,154]. Conventional fibre-fed spectrographs use MMFs to feed light to the spectrograph slit. However, the multimode core acts like a wide slit and so limits the spectrograph's resolution and compactness. It is possible to achieve diffraction-limited performance by reformatting the light from MM to SM with the photonic lantern [49]. This collects light at the lantern's MM port and uses the hexagonal array of SM cores in the MCF as a collection of spatially-separated entrance slits - a kind of unconventional compound pseudo-slit at the entrance to the spectrograph.

The image of this pseudo-slit on the spectrograph's detector is a 2D representation of data: one spatial, corresponding to the individual positions of the fibre cores at the entrance; and one spectral, corresponding to the dispersed spectrum of each core, Fig. 36. The images of the cores must be adequately sampled by the detector. To produce *N* non-overlapping spectra, one for each core, their images must be sufficiently separated on the detector along the spatial axis (ie, perpendicular to the dispersion axis). In a conventional 1D slit, individual fibres are normally placed in a row perpendicular to the dispersion axis, Fig. 36(a). In the TIGER configuration, a slight rotation between the dispersion axis of the spectrograph and the multicore array prevents the *N* spectra overlapping on the detector, Fig. 36(b).

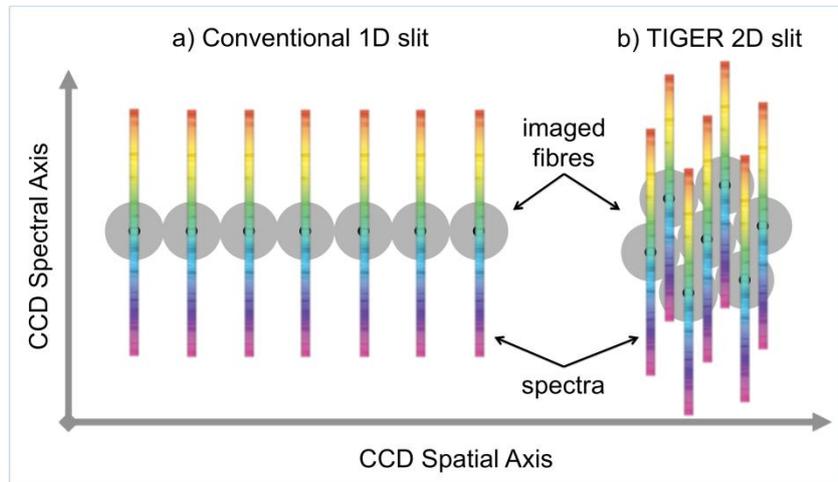

*Fig. 36. Schematic representation of the CCD sensor with two different fibre-based entrance configurations, (a) a conventional 1D slit and (b) a TIGER 2D slit.*

TIGER reduces the overall size of the optics needed by reformatting to a diffraction-limited state, while also reducing the overall size of the pseudo-slit. The width of a fibre slits is another important limit to the compactness of a fibre-fed spectrograph: the lenses and dispersion optics have to be many times larger than the entrance slit size to avoid off-axis aberrations, particularly at high resolution.

A photonic TIGER spectrograph can be realised by careful orientation of the multicore pseudo-slit. Fig. 37 shows a proof-of-concept spectrograph with 7 cores. The cores (with a mode field diameter of 10.5 μm at 1550 nm) were separated by 122 μm and the spectrograph was illuminated simultaneously with two laser sources, a 2 pm narrow-linewidth laser at 1549.95 nm and a 90 nm broadband SLD source centred on 1550 nm. The image of the cores and the spectra on the detector helps to assess the orientation of the cores against the dispersion axis of the spectrograph, Fig. 37. The lantern reformats the input MMF collection fibre into the 7 SM cores, and their detected individual spectra are add to obtain the total spectrum of the source in the bottom panel of Fig. 37. The large separation of the cores allows the slit to be oriented at many different angles from 9.5° and 35.5°. This Photonic TIGER spectrograph has already been demonstrated with 19-core type #3 lanterns [61]; and they are currently being developed as extreme-precision high-resolution spectrographs for astronomy and agriculture.

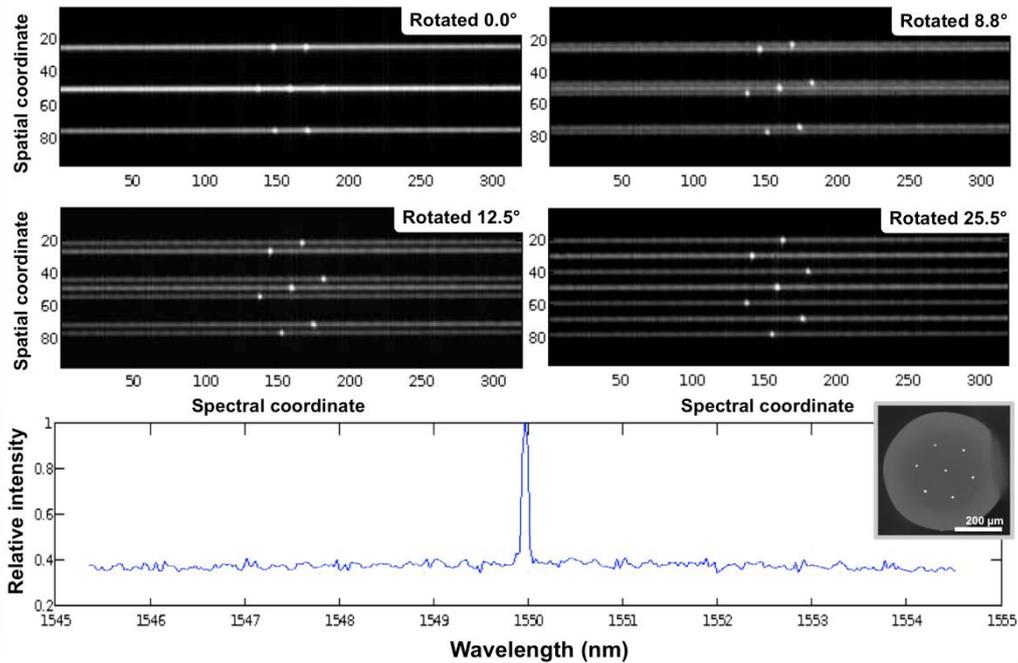

*Fig. 37. (four top panels) Spatial versus spectral coordinates detector image of the Photonic TIGER spectrograph for four pseudo-slit orientations, when fed light from a narrow laser and a broad SLD source simultaneously. (bottom panel) 1D processed spectrum from the 25.5° case showing the spectra of both sources. (bottom panel inset) Micrograph of the 7 core TIGER pseudo-slit.*

Another sense in which photonic lanterns can reformat MM light is directionally. Fig. 38 shows a few examples of devices made by connecting together the SM ports in various ways. None of these has been demonstrated experimentally yet, to our knowledge, though both type #2 (multi-fibre) and type #4 (ULI) methods would be suitable.

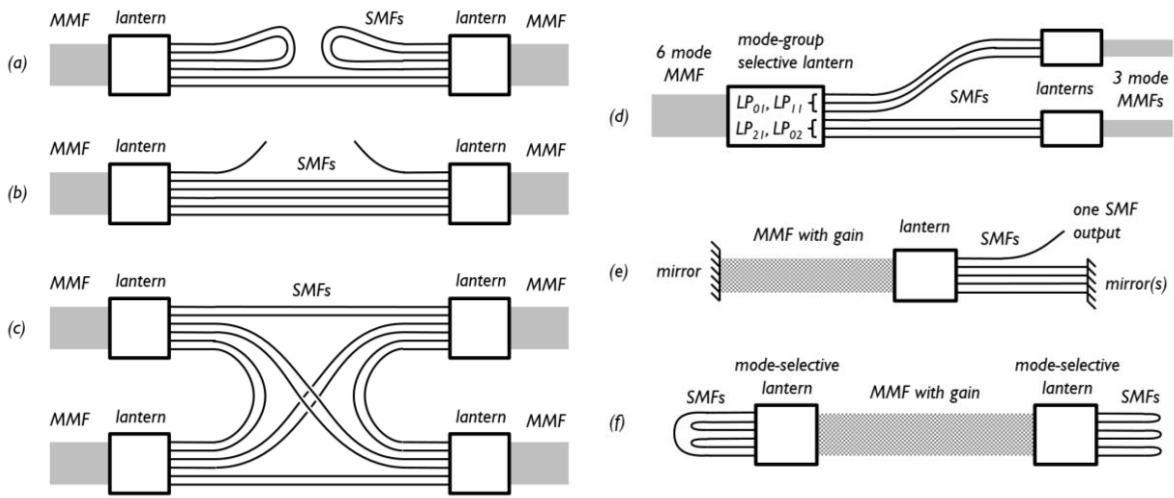

*Fig. 38. Directional devices based on photonic lanterns. (a) A partial reflector. (b) A tap. (c) An isotropic splitter. The number of MM ports is not restricted to the 4 shown, but for equal splitting it should be one greater than a factor of N. (d) A mode-selective MM splitter. (e) A multimode laser with a single-mode output. (f) A multipass cavity, each pass of the MM core being in a different mode. The usual laser paraphernalia (pump couplers, output couplers, mode-lockers etc) can be placed in the SM guides.*

### 4.4. Space division multiplexing

Early interest in photonic lanterns was based on their originally-envisaged applications in astronomical instrumentation [1-5,7-13,17-19,21-32,34,37,38] but recent attention has increasingly focused on their use in telecommunication [20,33,35,36,40,42,43,46-48,52,53,55,57,58,67,69-79] - an example of how new technologies find unexpected applications whose significance could potentially eclipse the original motivation. To increase capacity, the builders of optical telecommunication systems rely on various techniques to multiplex several channels of information onto a single fibre. The latest (and possibly final) degree of freedom to be exploited is the spatial distribution of the light, either between several separate cores in a multicore fibre or between several modes in a multimode (or "few-mode") fibre [75,155]. This technique is called space-division multiplexing (SDM). A multimode fibre with *N* modes in principle allows the capacity of the communication line to be multiplied by *N*. To implement SDM, spatial multiplexers are needed to couple different channels into the modes of the MMF at one end and then extract them at the other end. Direct mode multiplexing requiring each channel to be coupled to one unique mode is the simplest case to understand, but this is not necessary. The technique known as MIMO combines coherent detection with digital signal processing. This makes it sufficient for each channel to be imposed onto one unique combination of modes, provided it is orthogonal to the combinations of modes for all the other channels.

Spatial multiplexing can be carried out by photonic lanterns [20], the first experimental study of photonic lanterns in telecommunications being reported by Fontaine et al [33,35] in collaboration with one of us (SGL-S). Each of the lantern's SM ports delivers one channel of information (perhaps also multiplexed by wavelength or polarisation) that excites modes of the MM port, Fig. 39. Since the light patterns in the SM ports don't overlap each other, they are orthogonal, and unitary propagation along a loss-less lantern ensures that the mode distributions in the MM port are also orthogonal. Loss-less propagation in the lantern, including the avoidance of the symmetry losses discussed in Section 3.2.3, is important to prevent the loss of data [36].

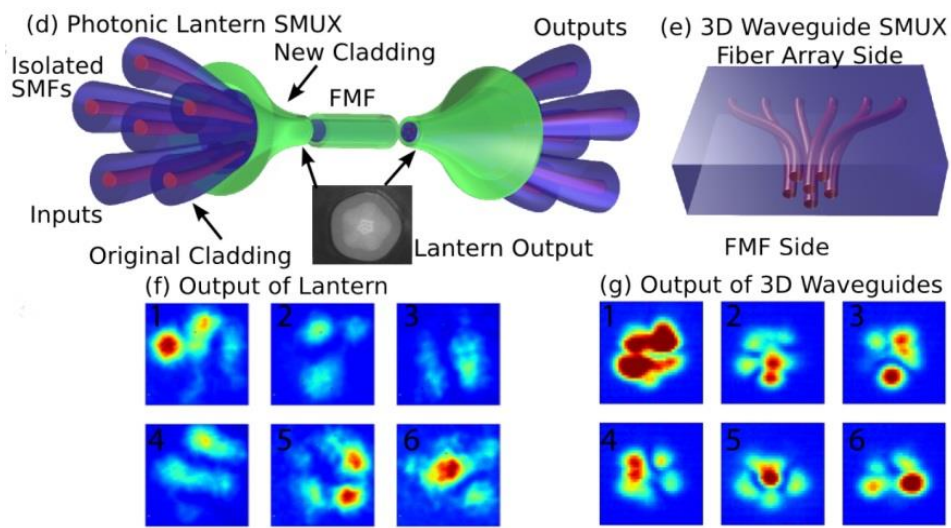

*Fig. 39. Spatial multiplexing using (left) type #2 and (right) type #4 photonic lanterns with N = 6 identical SM cores. In each case, upper pictures are schematics and lower pictures are output patterns in the MMF port (marked "FMF", meaning "few mode fibre"). The outputs are not pure modes but they are orthogonal combinations of the modes, so that information can be extracted by MIMO. Reprinted with permission from N. K. Fontaine et al OFC paper OTh1B.3 (2013) [40]. Copyright 2013, Optical Society of America.*

The ideas of using multimode or multicore fibres for SDM can also be combined by using few-mode MCFs, where each core of the MCF supports mode than one mode. Coupling into and out of such fibres is an area where ULI devices come into their own, recent examples of which consist of multiple photonic lanterns for coupling to each few-mode core of a 19 core MCF [57], Fig. 40. Such devices have also recently been used to demonstrate record data transmission rates of 200 Tbit/s over 1 km of fibre [74].

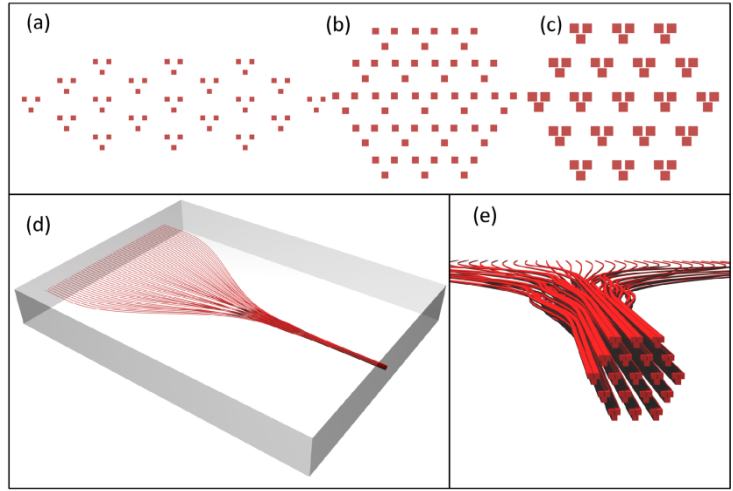

*Fig. 40. Schematic diagram of an interconnect device developed for few-mode MCF input & output coupling applications. a) The first intermediate plane. b) Second intermediate plane. c) Final core configuration. d) 3D schematic of waveguide paths. e) 3D schematic showing the spatial multiplexer output facet. Reprinted from [57].*

We note that some authors use the term "spot coupler" interchangeably with "photonic lantern", particularly when referring to ULI fabricated devices. It seems to us that the only difference between the ULI spot coupler and a ULI lantern is the degree to which widely-separated SM cores are brought together, to form either an MM core or a closer array of SM cores with some coupling between them. The devices of [50] and [57] could therefore be regarded either as spot couplers or photonic lanterns. However, if the cores are not brought together close enough to form supermodes then the device is not a photonic lantern but just a 3D fan-out [24,119]. Of course such devices do have applications in SDM, eg for interfacing to multicore fibres and for coupling to few-mode fibres using an aperture sampled approach [156], but they are outside the scope of this review.

The complexity of the digital processing required for MIMO scales with the differential group delay among each channel's modes, so it is advantageous to ensure that light from a given input is coupled only to one mode group rather than to all the available modes. Suitable mode-group-selective lanterns of type #2 (multi-fibre), in which the the necessary selectivity is provided by the use of sets of SMFs with dissimilar diameters, have been demonstrated for $N = 3$ with the $LP_{01}$ and $LP_{11}$ modes [48,53,67,72,73,78] and $N = 6$ with the $LP_{21}$ and $LP_{02}$ modes as well [67,76,77]. The rapidly-increasing pace of interest in mode-group-selectivity suggests that this may become a key real-world application for photonic lanterns.

Selectivity can be taken a step further to make direct mode multiplexers: the conceptually-simple case where each SM input excites just one mode of the MM output and MIMO is not necessary if the MMF's core is asymmetric [46,47,52]. This requires all of the SM cores to be dissimilar to each other. Then, in an adiabatic transition, light in the SMF of *n*-th greatest propagation constant $\beta$ excites the mode of *n*-th greatest $\beta$ in the output MMF [46,47,52,87,127,157-162]. Fig. 41 shows an example for $N = 3$ spatial modes. In that case, the initial fibres (and hence the lantern's SM ports) were identical but made locally dissimilar by pre-tapering to different extents before insertion in the F-doped capillary [47,52]. Similar behaviour was seen in an equivalent PCF device, an example of a type #5 lantern [46,52]. Mode-selective photonic lantern are also possible using ULI to fabricate, simply by changing parameters such as substrate translation speed and laser pulse energy for each SM core from which the lantern is constructed. This approach was used very recently to make a mode-selective tapered velocity coupler [126].

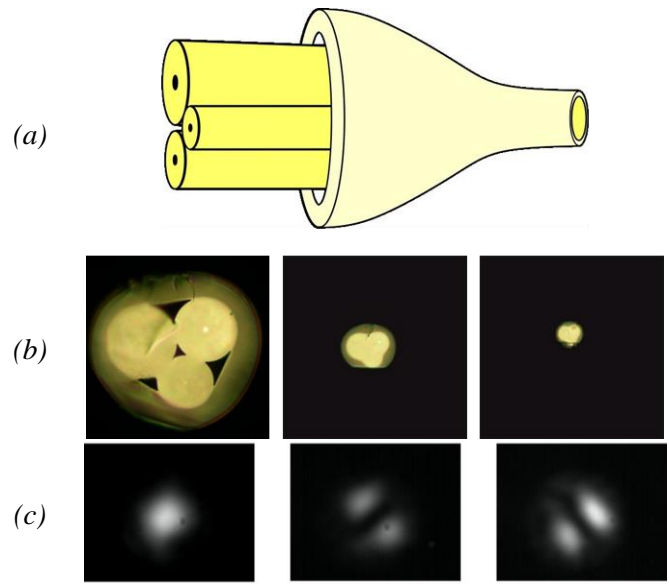

*Fig. 41. (a) Schematic mode multiplexer made from three locally-dissimilar SMFs. (b) cross-sections (to the same scale) along the taper transition, showing three locally-dissimilar SMFs merging to form an MMF core inside an F-doped cladding. (c) Output near-field patterns from the MMF port when each SMF (in decreasing order of diameter, left to right) is illuminated by 1550 nm light. The patterns match those expected for pure $LP_{01}$ and $LP_{11}$ modes. Reprinted from [52].*

Controlled excitation of pure modes in an MMF is also possible in principle by exciting the SM end of a non-mode-selective lantern (ie, not one of the lanterns described in the previous paragraph) with the appropriate amplitude and phase distribution. This would be illustrated by Fig. 1 or Fig. 17(a) in reverse. It has been demonstrated in a lantern-like silicon channel waveguide [129], though we are not aware of any experiments with fibre lanterns.

## 5. Why are they called "photonic lanterns"?

The waveguide structure that is the subject of this paper is now universally known as the photonic lantern. However, the origins of the name are not obvious [45]. In ordinary English a lantern is a source of light, yet photonic lanterns produce no light of their own. In fact the terminology has its origins in the schematic diagram of an MM-SM-MM lantern pair shown in Fig. 42(a) and reported in [3]. When the diagram was first drafted it reminded co-author J. Bland-Hawthorn of a paper lantern, Fig. 42(b). This became an informal name for the device, and it was used explicitly in subsequent papers [5,7,8,10]. Now the term "photonic lantern" is associated with the individual multicore SM-MM transition rather than the MM-SM-MM pair - further obscuring its origin!

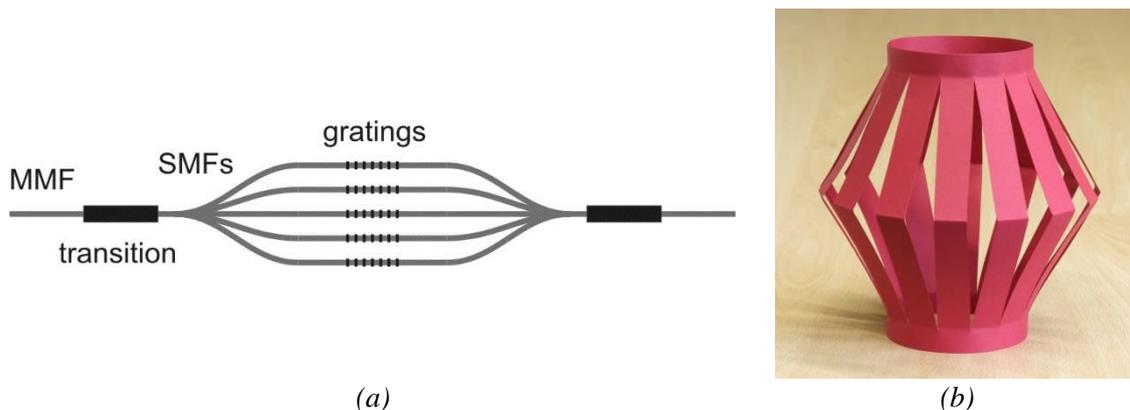

*(a)*                                             *(b)*

*Fig. 42. (a) The original schematic of an MM-SM-MM lantern pair [3], copied from Fig. 5. The gratings are incidental. (b) Photograph of a simple paper lantern.*

It is worth considering here what should be called a photonic lantern. The non-mode-selective MM-SM-MM lantern pair of [1,3] and Fig. 42(a), and its use to obtain SM performance in an MM device, were original to those papers. However, individual multicore SM-MM transitions like those that are now called photonic lanterns have a rather longer history. For example, the multiport fused couplers reported ~25 years ago by Mortimore and Arkwright [97-100], Fig. 9(c), looks just like an SM-MM-SM lantern pair - the structure of Fig. 42(a) but inside out. Indeed, cleaving the coupler at its waist (eg to produce Fig. 9(c)) would seem to yield two type #2 lanterns.

So, what is it about the more-recent photonic lanterns that justifies the new name? One view is that the key distinguishing features making a photonic lantern a photonic lantern are one or more of:
- the use of the multimode waveguide as an input and/or output port, not just an internal feature;
- at least three single-mode ports ($N > 2$), to exclude variants of the common $2 \times 2$ fused coupler;
- some consideration of mode-number matching in the design and operation of the device (if only to attain the "light bucket" condition, Fig. 20); and
- applications that are indifferent to how the light is distributed between the SM cores and/or the MM modes.

This last criterion is however inconsistent with the accepted status, as photonic lanterns, of the selective multiplexers described in Section 4.4.

Alternatively, one could quite consistently say that *any* multicore-to-multimode waveguide transition [81,85-87,97-100,127-129,151,160,161] is a photonic lantern, including half a multiport coupler, and that the terminology has merely evolved. Nevertheless we have focused on the more-recent work in this paper.

*6. Conclusions*

In this paper we have extensively reviewed the physics, fabrication and applications of photonic lanterns. Since they perform a relatively basic optical function - to interface between one multimode core and several single-mode cores - we look forward to the rise of new applications as fabrication techniques and performance improves.

*Acknowledgements*
Over the years our studies of photonic lanterns and allied subjects have received funding and other material support from various sources. At Bath we thank the Anglo-Australian Observatory, the Leibniz-Institut für Astrophysik Potsdam, the EPSRC and the Leverhulme Trust. At Heriot-Watt, RRT gratefully acknowledges funding from the STFC in the form of an STFC Advanced Fellowship (ST/H005595/1). For current funding, both institutions acknowledge the European Union via OPTICON Research Infrastructure for Optical/IR Astronomy (EU-FP7 226604) and the STFC via the Project Research and Development (STFC-PRD) scheme (ST/K00235X/1). We also thank the many individuals with whom we have worked on and discussed photonic lanterns over the years. We particularly acknowledge the exceptional contribution of Joss Bland-Hawthorn, who introduced us to astrophotonics and first asked whether SM action could be achieved in MMF, and with whom we have had many stimulating discussions and collaborations since.

*References*
[1] S. G. Leon-Saval, T. A. Birks, J. Bland-Hawthorn, M. Englund, "Single-mode performance in multimode fibre devices," Optical Fiber Communication Conference, Anaheim California (2005), paper PDP25.
[2] T. A. Birks, S. G. Leon-Saval, J. Bland-Hawthorn, M. Englund, "Adiabaticity in multicore fibre transitions," ACOFT Workshop on Optical Waveguide Theory and Numerical Modelling, Sydney Australia, 19 (2005).
[3] S. G. Leon-Saval, T. A. Birks, J. Bland-Hawthorn, M. Englund, "Multimode fiber devices with single-mode performance," Opt. Lett. **30**, 2545-2547 (2005).
[4] R. R. Thomson, A. K. Kar, J. Allington-Smith, "Ultrafast laser inscription: an enabling technology for astrophotonics," Opt. Express **17**, 1963-1969 (2009).

[5] D. Noordegraaf, P. M. W. Skovgaard, M. D. Nielsen, J. Bland-Hawthorn, "Efficient multi-mode to single-mode coupling in a photonic lantern," Opt. Express **17**, 1988-1994 (2009).

[6] Y. Shamir, Y. Sintov, M. Shtair, "Incoherent beam combining of multiple single-mode fiber lasers," Proc. SPIE 7580, 75801R (2010).

[7] D. Noordegraaf, P. M. W. Skovgaard, M. D. Maack, J. Bland-Hawthorn, R. Haynes, J. Lægsgaard, "Efficient multi-mode to single-mode conversion in a 61 port photonic lantern," Proc. SPIE 7580, 75802D (2010).

[8] D. Noordegraaf, P. M. W. Skovgaard, M. D. Maack, J. Bland-Hawthorn, R. Haynes, J. Lægsgaard, " Multi-mode to single-mode conversion in a 61 port photonic lantern," Opt. Express **18**, 4673-4678 (2010).

[9] S. G. Leon-Saval, A. Argyros, J. Bland-Hawthorn, "Photonic lanterns: a study of light propagation in multimode to single-mode converters," Opt. Express **18**, 8430-8439 (2010).

[10] J. Bland-Hawthorn, J. Lawrence, G. Robertson, S. Campbell, B. Pope, C. Betters, S. Leon-Saval, T. Birks, R. Haynes, N. Cvetojevic, N. Jovanovic, "PIMMS: photonic integrated multimode microspectrograph," Proc. SPIE 7735, 77350N (2010).

[11] S. C. Ellis, J. Lawrence, J. Bland-Hawthorn, J. Bryant, R. Haynes, A. Horton, S. Lee, S. Leon-Saval, H.-G. Löhmannsröben, J. Mladcnoff, J. O'Byrne, M. Rambold, M. Roth, C. Trinh, "GNOSIS: an OH suppression unit for near-infrared spectrographs," Proc. SPIE 7735, 773516 (2010).

[12] T. A. Birks, A. Díez, J. L. Cruz, S. G. Leon-Saval, D. F. Murphy, "Fibres are looking up: optical fibre transition structures in astrophotonics," Frontiers in Optics, Rochester New York (2010), paper FTuU1.

[13] R. R. Thomson, G. Brown, A. K. Kar, T. A. Birks, J. Bland-Hawthorn, "An integrated fan-out device for astrophotonics," Frontiers in Optics, Rochester New York (2010) paper PDPA3.

[14] Y. Shamir, Y. Sintov, M. Shtaif, "Beam quality analysis and optimization in an adiabatic low mode tapered fiber beam combiner," J. Opt. Soc. Am. B **27**, 2669-2676 (2010).

[15] D. Noordegraaf, M. D. Maack, P. M. W. Skovgaard, J. Johansen, F. Becker, S. Belke, M. Blomqvist, J. Laegsgaard, "All-fiber 7x1 signal combiner for incoherent laser beam combining," Proc. SPIE 7914, 79142L (2011).

[16] N. Mothe, P. Di Bin, "Multichannel microwave photonics signals summation device," Photon. Technol. Lett. **23**, 140-142 (2011).

[17] R. R. Thomson, T. A. Birks, S. G. Leon-Saval, A. K. Kar, J. Bland-Hawthorn, "Ultrafast laser inscription of an integrated photonic lantern," Opt. Express **19,** 5698-5705 (2011).

[18] T. A. Birks, B. J. Mangan, A. Díez, J. L. Cruz, S. G. Leon-Saval, J. Bland-Hawthorn, D. F. Murphy, "Multicore optical fibres for astrophotonics," European Conference on Lasers and Electro-Optics, Munich Germany (2011), paper JSIII2.1

[19] R. R. Thomson, T. A. Birks, S. G. Leon-Saval, A. K. Kar, J. Bland-Hawthorn, "Ultrafast laser inscription of an integrated multimode-to-single-modes waveguide transition for astrophotonics", European Conference on Lasers and Electro-Optics, Munich Germany (2011), paper JSIII2.2.

[20] H. Bülow, H. Al-Hashimi, and B. Schmauss, "Coherent multimode-fiber MIMO transmission with spatial constellation modulation," European Conference on Optical Communication, Geneva Switzerland (2011) paper Tu.5.B.3.

[21] J. Bland-Hawthorn, S. C. Ellis, S. G. Leon-Saval, R. Haynes, M. M. Roth, H.-G. Löhmannsröben, A. J. Horton, J.-G. Cuby, T. A. Birks, J. S. Lawrence, P. Gillingham, S. D. Ryder, C. Trinh, "A complex multi-notch astronomical filter to suppress the bright infrared sky," Nature Commun. **2**, 581 (2011).

[22] D. Noordegraaf, P. M. W. Skovgaard, R. H. Sandberg, M. D. Maack, J. Bland-Hawthorn, J. S. Lawrence, J. Lægsgaard, "Nineteen-port photonic lantern with multimode delivery fiber," Opt. Lett. **37,** 452-454 (2012).

[23] J. Bland-Hawthorn, P. Kern, "Molding the flow of light: photonics in astronomy," Phys. Today **65**(5), 31-37 (2012).

[24] R. R. Thomson, R. J. Harris, T. A. Birks, G. Brown, J. Allington-Smith, J. Bland-Hawthorn, "Ultrafast laser inscription of a 121-waveguide fan-out for astrophotonics," Opt. Lett. **37**, 2331-2333 (2012).

[25] T. A. Birks, B. J. Mangan, A. Díez, J.-L. Cruz, D. F. Murphy, ""Photonic lantern" spectral filters in multi-core fibre," Opt. Express **20**, 13996-14008 (2012).

[26] S. C. Ellis, J. Bland-Hawthorn, J. Lawrence, A. J. Horton, C. Trinh, S. G. Leon-Saval, K. Shortridge, J. Bryant, S. Case, M. Colless, W. Couch, K. Freeman, L. Gers, K. Glazebrook, R. Haynes, S. Lee, H.-G. Löhmannsröben, J. O'Byrne, S. Miziarski, M. Roth, B. Schmidt, C. G. Tinney, J. Zheng, "Suppression of the near-infrared OH night sky lines with fibre Bragg gratings – first results, " Mon. Not. R. Astron. Soc. **425**, 1682-1695 (2012).

[27] N. Jovanovic, I. Spaleniak, S. Gross, M. Ireland, J. S. Lawrence, C. Miese, A. Fuerbach, M. J. Withford, "Integrated photonic building blocks for next-generation astronomical instrumentation I: the multimode waveguide," Opt. Express **20,** 17029-17043 (2012).

[28] C. Trinh, S. C. Ellis, J. S. Lawrence, A. J. Horton, J. Bland-Hawthorn, S. G. Leon-Saval, J. Bryant, S. Case, M. Colless, W. Couch, K. Freeman, L. Gers, K. Glazebrook, R. Haynes, S. Lee, H.-G. Löhmannsröben, S. Miziarski, J. O'Byrne, M. Rambold, M. N. Roth, B. Schmidt, K. Shortridge, S. Smedley, C. G. Tinney, P. Xavier, J. Zheng, "GNOSIS: a novel near-infrared OH suppression unit at the AAT," Proc. SPIE 8446, 84463J (2012).

[29] W. Sun, H. Yu, Q. Yan, F. Tian, X. Liu, Y. Jiang, Z. Huang, "Light propagation in a fibre-brush-shape converter," Proc. SPIE 8450, 845014 (2012).

[30] I. Spaleniak, N. Jovanovic, S. Gross, M. Ireland, J. Lawrence, M. Withford, "Enabling photonic technologies for seeing-limited telescopes: fabrication of integrated photonic lanterns on a chip," Proc. SPIE 8450, 845015 (2012).

[31] A. Horton, S. Ellis, J. Lawrence, J. Bland-Hawthorn, "PRAXIS - a low background NIR spectrograph for fibre Bragg grating OH suppression," Proc. SPIE 8450, 84501V (2012).

[32] J.-C. Olaya, K. Ehrlich, D. M. Haynes, R. Haynes, S. G. Leon-Saval, D. Schirdewahn, "Multimode to single-mode converters: new results on 1-to-61 photonic lanterns," Proc. SPIE 8450, 84503K (2012).

[33] N. K. Fontaine, R. R. Ryf, S. G. Leon-Saval, J. Bland-Hawthorn, "Evaluation of photonic lanterns for lossless mode-multiplexing," European Conference on Optical Communication, Amsterdam Netherlands (2012), paper Th.2.D.6.

[34] J.-C. Olaya, S. G. Leon-Saval, D. Schirdewahn, K. Ehrlich, D. M. Haynes, R. Haynes, "1:61 photonic lanterns for astrophotometry: a performance study," Mon. Not. R. Astron. Soc. **427**, 1194-1208 (2012).

[35] R. Ryf, N. K. Fontaine, M. A. Mestre, S. Randel, X. Palou, C. Bolle, A. H. Gnauck, S. Chandrasekhar, X. Liu, B. Guan, R. Essiambre, P. J. Winzer, S. Leon-Saval, J. Bland-Hawthorn, R. Delbue, P. Pupalaikis, A. Sureka, Y. Sun, L. Grüner-Nielsen, R. V. Jensen, R. Lingle, "12 × 12 MIMO transmission over 130-km few-mode fiber," Frontiers in Optics, Rochester New York (2012), paper FW6C.4.

[36] N. K. Fontaine, R. Ryf, J. Bland-Hawthorn, S. G. Leon-Saval, "Geometric requirements for photonic lanterns in space division multiplexing," Opt. Express **20**, 27123-27132 (2012).

[37] T. A. Birks, B. J. Mangan, A. Díez, J.-L. Cruz, D. F. Murphy, "Multi-core fibre photonic lanterns," Australian Conference on Optical Fibre Technology, Sydney Australia (2012), paper 3G-1.

[38] R. R. Thomson, T. A. Birks, S. G. Leon-Saval, J. Bland-Hawthorn, P. S. Salter, R. D. Simmonds, M. Booth, "Ultrafast laser inscription of integrated 'photonic lanterns'," Australian Conference on Optical Fibre Technology, Sydney Australia (2012), paper 8G-1.

[39] C. Q. Trinh, S. C. Ellis, J. Bland-Hawthorn, J. S. Lawrence, A. J. Horton, S. G. Leon-Saval, K. Shortridge, J. Bryant, S. Case, M. Colless, W. Couch, K. Freeman, H.-G. Loehmannsroeben, L. Gers, K. Glazebrook, R. Haynes, S. Lee, J. O'Byrne, S. Miziarski, M. M. Roth, B. Schmidt, C. G. Tinney, J. Zheng, "GNOSIS: the first instrument to use fiber Bragg gratings for OH suppression, "Astron. J. **145**:51 (2013).

[40] N. K. Fontaine, "Devices and components for space-division multiplexing in few-mode fibers," Optical Fiber Communication Conference, Anaheim California (2013), paper OTh1B.3.

[41] C. Q. Trinh, S. C. Ellis, J. Bland-Hawthorn, A. J. Horton, J. S. Lawrence, S. G. Leon-Saval, "The nature of the near-infrared interline sky background using fibre Bragg grating OH suppression," Mon. Not. Roy. Astron. Soc. **432**, 3262-3277 (2013).

[42] N. K. Fontaine, R. Ryf, "Characterization of mode-dependent loss of laser inscribed photonic lanterns for space division multiplexing systems," Optoelectronics and Communications Conference, Kyoto Japan (2013), paper MR2-2.

[43] N. K. Fontaine, "Photonic lantern spatial multiplexers in space-division multiplexing," IEEE Photonics Soc. Summer Top. Meeting Ser. 97-98 (2013) paper TuC2.1.

[44] I. Ozdur, P. Toliver, A. Agarwal, T. K. Woodward, "Free-space to single-mode collection efficiency enhancement using photonic lanterns," Opt. Lett. **38**, 3554-3557 (2013).

[45] T. A. Birks, "Photonic lanterns," presented at the European Conference on Optical Communication Workshop WS3 on Integration of Optical Devices for SDM Transmission, London UK (2013), http://www.ecoc2013.org/docs/tim-birks.pdf

[46] S. Yerolatsitis, T. A. Birks, "Three-mode multiplexer in photonic crystal fibre," European Conference on Optical Communication, London UK (2013), paper Mo.4.A.4.

[47] S. Yerolatsitis, T. A. Birks, "Tapered mode multiplexers based on standard single-mode fibre," European Conference on Optical Communication, London UK (2013), paper PD1.C.1.

[48] N. K. Fontaine, S. G. Leon-Saval, R. Ryf, J. R. Salazar Gil, B. Ercan, J. Bland-Hawthorn, "Mode-selective dissimilar fiber photonic-lantern spatial multiplexers for few-mode fiber," European Conference on Optical Communication, London UK (2013), paper PD1.C.3.

[49] C. H. Betters, S. G. Leon-Saval, J. G. Robertson, J. Bland-Hawthorn, "Beating the classical limit: A diffraction-limited spectrograph for an arbitrary input beam," Opt. Express **21**, 26103-26112 (2013).

[50] I. Spaleniak, N. Jovanovic, S. Gross, M. J. Ireland, J. S. Lawrence, M. J. Withford, "Integrated photonic building blocks for next-generation astronomical instrumentation II: the multimode to single mode transition," Opt. Express **21**, 27197-27208 (2013).

[51] S. G. Leon-Saval, A. Argyros, J. Bland-Hawthorn, "Photonic lanterns," Nanophoton. **2**, 429-440 (2013).

[52] S. Yerolatsitis, I. Gris-Sánchez, T. A. Birks, "Adiabatically-tapered fiber mode multiplexers," Opt. Express **22**, 608-617 (2014).

[53] S. G. Leon-Saval, N. K. Fontaine, J. R. Salazar-Gil, B. Ercan, R. Ryf, J. Bland-Hawthorn, "Mode-selective photonic lanterns for space-division multiplexing," Opt. Express **22**, 1036-1044 (2014).

[54] I. Spaleniak, S. Gross, N. Jovanovic, R. J. Williams, J. S. Lawrence, M. J. Ireland, M. J. Withford, "Multiband processing of multimode light: combining 3D photonic lanterns with waveguide Bragg gratings," Laser Photon. Rev. **8**, L1-L5 (2014).

[55] J. Carpenter, S. G. Leon-Saval, J. R. Salazar-Gil, J. Bland-Hawthorn, G. Baxter, L. Stewart, S. Frisken, M. A. F. Roelens, B. J. Eggleton, J. Schröder, "1×11 few-mode fiber wavelength selective switch using photonic lanterns," Opt. Express **22**, 2216-2221 (2014).

[56] R. J. Harris, D. G. MacLachlan, D. Choudhury, T. J. Morris, E. Gendron, A. G. Basden, G. Brown, J. R. Allington-Smith, R. R. Thomson, "Photonic spatial reformatting of stellar light for diffraction-limited spectroscopy," arXiv:1402.2547 [astro-ph.IM] (2014); accepted for publication in Mon. Not. Roy. Astron. Soc., Feb. 2015).

[57] P. Mitchell, G. Brown, R. R. Thomson, N. Psaila, A. Kar, "57 Channel (19×3) Spatial multiplexer fabricated using direct laser inscription," Optical Fiber Communication Conference, San Francisco California (2014), paper M3K.5.

[58] I. Giles, R. Chen, V. Garcia-Munoz, "Fiber based multiplexing and demultiplexing devices for few mode fiber space division multiplexed communications," Optical Fiber Communication Conference, San Francisco California (2014), paper Tu3D.1.

[59] T. A. Birks, I. Gris-Sánchez, S. Yerolatsitis, "The photonic lantern," Conference on Lasers and Electro-Optics, San Jose California (2014), paper SM2N.3.

[60] D. M. Haynes, I. Gris-Sánchez, K. Ehrlich, T. A. Birks, R. Haynes, "New multicore low mode noise scrambling fiber for applications in high-resolution spectroscopy", Proc. SPIE 9151, 915155 (2014).

[61] C. H. Betters, S. G. Leon-Saval, J. Bland-Hawthorn, S. N. Richards, T. A. Birks, I. Gris-Sanchez, " PIMMS échelle: the next generation of compact diffraction limited spectrographs for arbitrary input beams," Proc. SPIE 9147, 91471I (2014).

[62] A. Arriola, D. Choudhury, R. R. Thomson, "New generation of photonic lanterns for mid-IR astronomy," Proc. SPIE 9151, 915116 (2014).

[63] D. G. MacLachlan, R. Harris, D. Choudhury, A. Arriola, G. Brown, J. Allington-Smith, R. R. Thomson, "Development of integrated "photonic-dicers" for reformatting the point-spread-function of a telescope," Proc. SPIE 9151, 91511W (2014).

[64] A. Horton, R. Content, S. Ellis, J. Lawrence, "Photonic lantern behaviour and implications for instrument design," Proc. SPIE 9151, 915122 (2014).

[65] W. Sun, Q. Yan, Y. Bi, H. Yu, X. Liu, J. Xue, H. Tian, Y. Liu, "Photonic lantern with cladding-removable fibers," Proc. SPIE 9151, 91514C (2014).

[66] R. Content, J. Bland-Hawthorn, S. Ellis, L. Gers, R. Haynes, A. Horton, J. Lawrence, S. Leon-Saval, E. Lindley, S.-S. Min, K. Shortridge, N. Staszak, C. Trinh, P. Xavier, R. Zhelern, "PRAXIS: low thermal emission high efficiency OH suppressed fibre spectrograph," Proc. SPIE 9151, 91514W (2014).

[67] S. G. Leon-Saval, "Multimode photonics, optical transition devices for multimode control," Optoelectronics and Communications Conference, Melbourne Australia (2014), paper MO2C-5.

[68] H.-J. Yu, Q. Yan, Z.-J. Huang, H. Tian, Y. Jiang, Y.-J. Liu, J.-Z. Zhang, W.-M. Sun, "Photonic lantern with multimode fibers embedded," Res. Astron. Astrophys. **14**, 1046-1054 (2014).

[69] Y. Chen, A. Lobato, Y. Jung, H. Chen, R. V. Jensen, Y. Sun, L. Grüner-Nielsen, D. J. Richardson, V. A. J. Sleiffer, M. Kuschnerov, N. K. Fontaine, R. Ryf, I. P. Giles, R. Chen, V. Carcia-Munoz, A. M. J. Koonen, B. Lanki, N. Hanik, "41.6 Tb/s C-band SDM OFDM transmission through 12 spatial and polarization modes over 74.17 km few mode fiber," European Conference on Optical Communication, Cannes France (2014), paper Mo.3.3.3.

[70] R. G. H. van Uden, R. Amezcua Correa, E. Antonio-Lopez, F. M. Huijskens, G. Li, A. Schulzgen, H. de Waardt, A. M. J. Koonen, C. M. Okonkwo, "1 km hole-assisted few-mode multi-core fiber 32QAM WDM transmission," European Conference on Optical Communication, Cannes France (2014), paper Mo.3.3.4.

[71] H. Chen, N. K. Fontaine, R. Ryf, B. Guan, S. J. B. Yoo, A. M. J. Koonen, "A fully-packaged 3D-waveguide based dual-fiber spatial-multiplexer with up-tapered 6-mode fiber pigtails," European Conference on Optical Communication, Cannes France (2014), paper We.1.1.4.

[72] C. Xia, N. Chand, A. M. Velázquez-Benitez, X. Liu, J. E. A. Lopez, H. Wen, B. Zhu, F. Effenberger, R. Amezcua-Correa, G. Li, "Demonstration of World's first few-mode GPON," European Conference on Optical Communication, Cannes France (2014), paper PD.1.5.

[73] R. Ryf, N. K. Fontaine, B. Guan, B. Huang, M. Esmaeelpour, S. Randel, A. H. Gnauk, S. Chandrasekhar, A. Adamiecki, G. Raybon, R. W. Tkach, R. Shubochkin, Y. Sun, R. Lingle, "305-km combined wavelength and mode-multiplexed transmission over conventional graded-index multimode fibre," European Conference on Optical Communication, Cannes France (2014), paper PD.3.5.

[74] R. G. H. van Uden, R. Amezcua Correa, E. A. Lopez, F. M. Huijskens, C. Xia, G. Li, A. Schülzgen, H. de Waardt, A. M. J. Koonen, C. M. Okonkwo, "Ultra-high-density spatial division multiplexing with a few-mode multicore fibre," Nature Photon. **8**, 865-870 (2014).

[75] G. Li, N. Bai, N. Zhao, C. Xia, "Space-division multiplexing: the next frontier in optical communication," Adv. Opt. Photon. **6**, 413-487 (2014).

[76] B. Huang, N. K. Fontaine, R. Ryf, B. Guan, S. G. Leon-Saval, R. Shubochkin, Y. Sun, R. Lingle, G. Li, "All-fiber mode-group-selective photonic lantern using graded-index multimode fibers," Opt. Express **23**, 224-234 (2015).

[77] R. Ryf, N. K. Fontaine, H. Chen, B. Guan, B. Huang, M. Esmaeelpour, A. H. Gnauk, S. Randel, S. J. B. Yoo, A. M. J. Koonen, R. Shubochkin, Y. Sun, R. Lingle, "Mode-multiplexed transmission over conventional graded-index multimode fibers," Opt. Express **23**, 235-246 (2015).

[78] C. Xia, N. Chand, A. M. Velázquez-Benitez, Z. Yang, X. Liu, J. E. Antonio-Lopez, H. Wen, B. Zhu, N. Zhao, F. Effenberger, R. Amezcua-Correa, G. Li, "Time-division-multiplexed few-mode passive optical network," Opt. Express **23**, 1151-1158.

[79] http://optoscribe.com/products/3d-mmux-photonic-lantern/

[80] N. D. Psaila (Optoscribe), private communication.

[81] B. S. Kawasaki, K. O. Hill, R. G. Lamont, "Biconical-taper single-mode fiber coupler," Opt. Lett. **6**, 327-328 (1981).

[82] J. Bures, S. Lacroix, J. Lapierre, "Analyse d'un coupleur bidirectionnel à fibres optiques monomodes fusionnées," Appl. Opt. **22**, 1918-1922 (1983)

[83] S. Lacroix, F. Gonthier, J. Bures, "Modeling of symmetric 2×2 fused-fiber couplers," Appl. Opt. **33**, 8361-8369 (1994).

[84] W. K. Burns, M. Abebe, C. A. Villarruel, "Parabolic model for shape of fiber taper," Appl. Opt. **24**, 2753-2755 (1985).

[85]  M. N. McLandrich, "Core dopant profiles in weakly fused single-mode fibres," Electron. Lett. **24**, 8-10 (1988).
[86]  F. Bilodeau, K. O. Hill, S. Faucher, D. C. Johnson, "Low-loss highly overcoupled fused couplers: fabrication and sensitivity to external pressure," J. Lightwave Technol. **6**, 1476-1482 (1988).
[87]  T. A. Birks, D. O. Culverhouse, S. G. Farwell, P. St.J. Russell, "All-fiber polarizer based on a null taper coupler," Opt. Lett. **20**, 1371-1373 (1995).
[88]  C. D. Hussey, T. A. Birks, A. M. Wingfield, A. Niu, R. Kenny, L. Dong, "Rectangular mode transformers in tapered single-mode fibres," Electron. Lett. **25**, 1419-1420 (1989).
[89]  T. A. Birks, Y. W. Li, "The shape of fiber tapers," J. Lightwave Technol. **10**, 432-438 (1992).
[90]  A. Witkowska, K. Lai, S. G. Leon-Saval, W. J. Wadsworth, T. A. Birks, "All-fibre anamorphic core-shape transitions," Opt. Lett. **31**, 2672-2674 (2006).
[91]  R. Falciai, A. M. Scheggi, "Biconical fibers as mode and wavelength filters," Appl. Opt. **28**, 1309-1311 (1989).
[92]  T. Schwander, B. Schwaderer, H. Storm, "Coupling of lasers to single-mode fibres with high efficiency and low optical feedback," Electron. Lett. **21**, 287-289 (1985).
[93]  H. Yokota, E. Sugai, Y. Saaki, "Optical irradiation method for fiber coupler fabrications," Opt. Rev. **4**, 104-107 (1997).
[94]  J. D. Love, "Spot size, adiabaticity and diffraction in tapered fibres," Electron. Lett. **23**, 993-994 (1987).
[95]  S. G. Leon-Saval, T. A. Birks, N. Y. Joly, A. K. George, W. J. Wadsworth, G. Kakarantzas, P. St.J. Russell, "Splice-free interfacing of photonic crystal fibers," Opt. Lett. **30**, 1629-1631 (2005).
[96]  A. W. Snyder, J. D. Love, "Optical Waveguide Theory," Chapman and Hall, London (1983).
[97]  D. B. Mortimore, J. W. Arkwright, "Monolithic wavelength-flattened 1×7 single-mode fused coupler," Electron. Lett. **25**, 606-607 (1989).
[98]  D. B. Mortimore "Monolithic 4×4 single-mode fused coupler," Electron. Lett. **25**, 682-683 (1989).
[99]  J. W. Arkwright, D. B. Mortimore, R. M. Adnams, "Monolithic 1×19 single-mode fused fibre couplers," Electron. Lett. **27**, 737-738 (1991).
[100] D. B. Mortimore, J. W. Arkwright, "Monolithic wavelength-flattened 1×7 single-mode fused fibre couplers: theory, fabrication, and analysis," Appl. Opt. **30**, 650-659 (1991).
[101] E. Kapon, J. Katz, A. Yariv, "Supermode analysis of phase-locked arrays of semiconductor lasers," Opt. Lett. **10**, 125-127 (1984).
[102] F. Ladouceur, J. D. Love, "Multiport single-mode fibre splitters," Opt. Quantum Electron. **22**, 453-465 (1990).
[103] J. Bland-Hawthorn, M. Englund, G. Edvell, "New approach to atmospheric OH suppression using an aperiodic fibre Bragg grating," Opt. Express **12**, 5902-5909 (2004).
[104] T. Mizunami, T. V. Djambova, T. Niiho, S. Gupta, "Bragg gratings in multimode and few-mode optical fibers," J. Lightwave Technol. **18**, 230-235 (2000).
[105] P. St.J. Russell, "Photonic-crystal fibers," J. Lightwave Technol. **24**, 4729-4749 (2006).
[106] Unpublished experiments at the University of Bath by A. Witkowska and S. G. Leon-Saval (2006).
[107] K. M. Davis, K. Miura, N. Sugimoto, K. Hirao, "Writing waveguides in glass with a femtosecond laser," Opt. Lett. **21**, 1729-1731 (1996).
[108] E. N. Glezer, M. Milosavljevic, L. Huang, R. J. Finlay, T.-H. Her, J. P. Callan, E. Mazur, "Three-dimensional optical storage inside transparent materials," Opt. Lett. **21**, 2023-2025 (1996).
[109] J. W. Chan, T. Huser, S. Risbud, D. M. Krol, "Structural changes in fused silica after exposure to focused femtosecond laser pulses," Opt. Lett. **26**, 1726-1728 (2001).
[110] E. N. Glezer, E. Mazur, "Ultrafast-laser driven micro-explosions in transparent materials," Appl. Phys. Lett. **71**, 882-884 (1997).
[111] Y. Liu, M. Shimizu, B. Zhu, Y. Dai, B. Qian, J. Qiu, Y. Shimotsuma, K. Miura, K. Hirao, "Micromodification of element distribution in glass using femtosecond laser irradiation," Opt. Lett. **34**, 136-138 (2009).
[112] Y. Shimotsuma, P. G. Kazansky, J. Qiu, K. Hirao, "Self-organized nanogratings in glass irradiated by ultrashort light pulses," Phys. Rev. Lett. **91**, 247405 (2003).
[113] A. Marcinkevičius, S. Juodkazis, M. Watanabe, M. Miwa, S. Matsuo, H. Misawa, J. Nishii, "Femtosecond laser-assisted three-dimensional microfabrication in silica," Opt. Lett. **26**, 277-279 (2001).

[114] Y. Bellouard, E. Barthel, A. A. Said, M. Dugan, P. Bado, "Scanning thermal microscopy and Raman analysis of bulk fused silica exposed to low energy femtosecond laser pulses," Opt. Express **16**, 19520–19534 (2008).

[115] S. Zhou, W. Lei, J. Chen, J. Hao, H. Zeng, J. Qiu, "Laser-induced optical property changes inside Bi-doped glass," Photon. Technol. Lett. **21**, 386-388 (2009).

[116] S. Nolte, M. Will, J. Burghoff, A. Tünnermann, "Femtosecond waveguide writing: a new avenue to three-dimensional integrated optics," Appl. Phys. A **77**, 109-111 (2003).

[117] R. Osellame, N. Chiodo, G. Della Valle, S. Taccheo, R. Ramponi, G. Cerullo, A. Killi, U. Morgner, M. Lederer, D. Kopf, "Optical waveguide writing with a diode-pumped femtosecond oscillator," Opt. Lett. **29**, 1900-1902 (2004).

[118] S. Taccheo, G. Della Valle, R. Osellame, G. Cerullo, N. Chiodo, P. Laporta, O. Svelto, A. Killi, U. Morgner, M. Lederer, D. Kopf, "Er:Yb-doped waveguide laser fabricated by femtosecond laser pulses," Opt. Lett. **29**, 2626-2628 (2004).

[119] R. R. Thomson, H. T. Bookey, N. D. Psaila, A. Fender, S. Campbell, W. N. MacPherson, J. S. Barton, D. T. Reid, A. K. Kar, "Ultrafast-laser inscription of a three dimensional fan-out device for multicore fiber coupling applications," Opt. Express **15**, 11691-11697 (2007).

[120] S. J. Beecher, R. R. Thomson, G. Brown, A. S. Webb, J. Sahu, A. K. Kar, "Bragg grating waveguide array ultrafast laser inscribed into the cladding of a flat fiber", MATEC Web of Conferences **8**, 6001 (2013).

[121] J. R. Grenier, L. A. Fernandes, P. R. Herman, "Femtosecond laser writing of optical edge filters in fused silica optical waveguides," Opt. Express **21**, 4493-4502 (2013).

[122] A. A. Said, M. Dugan, P. Bado, Y. Bellouard, A. Scott, J. Mabesa, "Manufacturing by laser direct-write of three-dimensional devices containing optical and microfluidic networks," Proc. SPIE 5339, 194-204 (2004).

[123] Y. Nasu, M. Kohtoku, Y. Hibino, "Low-loss waveguides written with a femtosecond laser for flexible interconnection in a planar light-wave circuit," Opt. Lett. **30**, 723-725 (2005).

[124] A. Arriola, S. Gross, N. Jovanovic, N. Charles, P. G. Tuthill, S. M. Olaizola, A. Fuerbach, M. J. Withford, "Low bend loss waveguides enable compact, efficient 3D photonic chips," Opt. Express **21**, 2978-2986 (2013).

[125] T. Meany, S. Gross, N. Jovanovic, A. Arriola, M. J. Steel, M. J. Withford, "Towards low-loss lightwave circuits for non-classical optics at 800 nm and 1550 nm," Appl. Opt. A **114**, 113-118 (2014).

[126] S. Gross, N. Riesen, J. D. Love, M. J. Withford, "Three-dimensional ultra-broadband integrated tapered mode multiplexers," Laser Photon. Rev. **8**, L81–L85 (2014).

[127] G. J. Veldhuis, J. H. Berends, P. V. Lambeck, "Design and characterization of a mode-splitting Ψ-junction," J. Lightwave Technol. **14**, 1746-1752 (1996).

[128] B.-T. Lee, S.-Y. Shin, "Mode-order converter in a multimode waveguide," Opt. Lett. **28**, 1660-1662 (2003).

[129] J. B. Park, D.-M. Yeo, S.-Y. Shin, "Variable optical mode generator in a multimode waveguide," Photon. Technol. Lett. **18**, 2084-2086 (2006).

[130] D. H. McMahon, "Efficiency limitations imposed by thermodynamics on optical coupling in fiber-optic data links," J. Opt. Soc. Am **65**, 1479-1482 (1975).

[131] D. A. B. Miller, "Self-aligning universal beam coupler," Opt. Express **21**, 6360-6370 (2013).

[132] Our thinking on waveguide adaptive optics has benefited from discussions with D. A. B. Miller.

[133] B. E. A. Saleh, M. C. Teich, "Fundamentals of Photonics," John Wiley & Sons, New York (1991). We could only find Eq. (5b), as used in [5,8,10], in the caption of Fig. 8.1-8 of this reference. That figure shows $N(V)$ like our Fig. 19 but incorrectly calculated.

[134] C. A. Clayton, "The implications of image scrambling and focal ratio degradation in fibre optics on the design of astronomical instrumentation," Astron. Astrophys. **213**, 502-515 (1989).

[135] T. A. Birks, J. C. Knight, P. St.J. Russell, "Endlessly single-mode photonic crystal fiber," Opt. Lett. **22**, 961-963 (1997).

[136] W. Jin, Z. Wang, J. Ju, "Two-mode photonic crystal fibers," Opt. Express **13**, 2082-2088 (2005).

[137] R. F. Harrington, "Time-Harmonic Electromagnetic Fields," John Wiley & Sons, New York (2001).

[138] H.-J. Yoo, J. R. Hayes, E. G. Paek, A. Scherer, Y.-S. Kwon, "Array mode analysis of two-dimensional phased arrays of vertical cavity surface emitting lasers," J. Quantum. Electron. **26**, 1039-1051 (1990).
[139] N. W. Ashcroft, N. D. Mermin, "Solid State Physics," Harcourt Brace, Fort Worth (1976).
[140] M. E. Zoorob, M. D. B. Charlton, G. J. Parker, J. J. Baumberg, M. C. Netti, "Complete photonic bandgaps in 12-fold symmetric quasicrystals," Nature **404**, 740-743 (2000).
[141] S. F. Liew, H. Noh, J. Trevino, L. Dal Negro, H. Cao, "Localized photonic band edge modes and orbital angular momenta of light in a golden-angle spiral," Opt. Express **19**, 23631-23642 (2011).
[142] J. D. Love, W. M. Henry, "Quantifying loss minimisation in single-mode fibre tapers," Electron. Lett. **22**, 912-914 (1986).
[143] J. D. Love, W. M. Henry, W. J. Stewart, R. J. Black, S. Lacroix, F. Gonthier, "Tapered single-mode fibres and devices part 1: adiabaticity criteria," IEE Proc. Pt. J **138**, 343-354 (1991).
[144] A. W. Snyder, X.-H. Zheng, "Optical fibers of arbitrary cross sections," J. Opt. Soc. Am A **3**, 600-609 (1986).
[145] G. Meltz, W. W. Morey, W. H. Glenn, "Formation of Bragg gratings in optical fibers by a transverse holographic method," Opt. Lett. **14**, 823-825 (1989).
[146] J. Bland-Hawthorn, P. Kern, "Astrophotonics: a new era for astronomical instruments," Opt. Express **17**, 1880-1884 (2009).
[147] R. Haynes, T. A. Birks, J. Bland-Hawthorn, J. L. Cruz, A. Diez, S. C. Ellis, D. Haynes, R. G. Krämer, B. J. Mangan, S. Min, D. F. Murphy, S. Nolte, J. C. Olaya, J. U. Thomas, C. Q. Trinh, A. Tünnermann, C. Voigtländer, "Second generation OH suppression filters using multicore fibres," Proc. SPIE 8450, 845011 (2012).
[148] S.-S. Min, C. Trinh, S. Leon-Saval, N. Jovanovic, P. Gillingham, J. Bland-Hawthorn, J. Lawrence, T. A. Birks, M. M. Roth, R. Haynes, L. Fogarty, "Multi-core fiber Bragg grating developments for OH suppression," Proc. SPIE 8450, 84503L (2012).
[149] E. Lindley, S.-S. Min, S. Leon-Saval, N. Cvetojevic, N. Jovanovic, J. Bland-Hawthorn, J. Lawrence, I. Gris-Sanchez, T. Birks, R. Haynes, D. Haynes, "Core-to-core uniformity improvements in multi-core fiber Bragg gratings," Proc. SPIE 9151, 91515F (2014).
[150] E. Lindley, S.-S. Min, S. Leon-Saval, N. Cvetojevic, J. Lawrence, S. Ellis, J. Bland-Hawthorn, "Demonstration of uniform multicore fiber Bragg gratings," Opt. Express **22**, 31575-31581 (2014).
[151] V. A. Kozlov, J. Hernández-Cordero, T. F. Morse, "All-fiber coherent beam combining of fiber lasers," Opt. Lett. **24**, 1814-1816 (1999).
[152] M. J. Hackert, "Explanation of launch condition choice for GRIN multimode fiber attenuation and bandwidth measurements," J. Lightwave Technol. **10**, 125-129 (1992).
[153] D. Donlagic, "A low bending loss multimode fiber transmission system," Opt. Express **17**, 22081-22095 (2009).
[154] S. G. Leon-Saval, C. H. Betters, J. Bland-Hawthorn, "The Photonic TIGER: a multicore fiber-fed spectrograph", Proc. SPIE 8450, 84501K (2012).
[155] D. J. Richardson, J. M. Fini, L. E. Nelson, "Space-division multiplexing in optical fibres," Nature Photon. **7**, 354-362 (2013).
[156] M. Blau, D. M. Marom, "Optimization of spatial aperture-sampled mode multiplexer for a three-mode fiber," Photon. Technol. Lett. **24**, 2101-2104 (2012).
[157] W. K. Burns, A. F. Milton, "Mode conversion in planar-dielectric separating waveguides," J. Lightwave Technol. **11**, 32-39 (1975).
[158] E. Kapon, R. N. Thurston, "Multichannel waveguide junctions for guided-wave optics," Appl. Phys. Lett. **50**, 1710-1712 (1987).
[159] R. W. C. Vance, J. D. Love, "Asymmetric adiabatic multiprong for mode-multiplexed systems," Electron. Lett. **29**, 2134-2136 (1993).
[160] T. A. Birks, S. G. Farwell, P. St.J. Russell, C. N. Pannell, "Four-port fiber frequency shifter with a null taper coupler," Opt. Lett. **19**, 1964-1966 (1994); erratum Opt. Lett. **21**, 231 (1996).
[161] K. Lai, S. G. Leon-Saval, A. Witkowska, W. J. Wadsworth, T. A. Birks, "Wavelength-independent all-fiber mode converters," Opt. Lett. **32**, 328-330 (2007).
[162] N. Riesen, J. D. Love, J. W. Arkwright, "Few-mode elliptical core fiber data transmission," Photon. Technol. Lett. **24**, 344-346 (2012).